\documentclass[12pt,a4paper]{article}
\pdfoutput=1

\usepackage{amssymb}
\usepackage{amsmath}
\usepackage{amsfonts}
\usepackage{amsthm}
\usepackage{setspace}
\usepackage{color}
\usepackage[pdftex]{graphicx}
\usepackage{authblk}
\usepackage{subfigure}
\usepackage{float}
\usepackage[utf8]{inputenc}

\numberwithin{equation}{section}

\newcommand{\BbbR}{\mathbb{R}}
\newcommand{\BbbZ}{\mathbb{Z}}
\newcommand{\BbbC}{\mathbb{C}}

\DeclareMathOperator{\diag}{diag}
\DeclareMathOperator{\Tr}{Tr}
\DeclareMathOperator{\erfc}{erfc} 

\theoremstyle{plain}
\newtheorem{thm}{Theorem}

\title{Unruh-DeWitt detector's response to fermions\\ 
in flat spacetimes}

\author{Jorma Louko\thanks{jorma.louko@nottingham.ac.uk}\ }
\author{Vladimir Toussaint\thanks{pmxvt2@nottingham.ac.uk}}
\affil{School of Mathematical Sciences\\ 
University of Nottingham\\ 
Nottingham NG7 2RD\\ 
UK}

\date{\small August 2016\\[1ex]
Published in Phys.\ Rev.\ D {\bf94}, 064027 (2016)}

\begin{document}

\maketitle
\begin{abstract}
We examine an Unruh-DeWitt particle detector that is coupled linearly
to the scalar density of a massless Dirac field in 
Minkowski spacetimes of dimension $d\ge2$ 
and on the static Minkowski cylinder 
in spacetime dimension two, allowing the detector's motion to 
remain arbitrary and working to leading order in perturbation theory. 
In $d$-dimensional Minkowski, with the field in the usual Fock vacuum, 
we show that the detector's response is 
identical to that of a detector coupled 
linearly to a massless scalar field in $2d$-dimensional Minkowski. 
In the special case of uniform linear acceleration, 
the detector's response 
hence exhibits the Unruh effect with a Planckian factor 
in both even and odd dimensions, 
in contrast to the Rindler power spectrum of the Dirac 
field, which has a Planckian factor for odd $d$ but a 
Fermi-Dirac factor for even~$d$. 
On the two-dimensional cylinder, we set the oscillator 
modes in the usual Fock vacuum 
but allow an arbitrary state for the zero mode of the periodic spinor. 
We show that the 
detector's response distinguishes the periodic and antiperiodic spin
structures, and the zero mode of the periodic spinor contributes to the 
response by a state-dependent but well defined amount. 
Explicit analytic and numerical results on the
cylinder are obtained for inertial and uniformly accelerated
trajectories, recovering the $d=2$ Minkowski results in the limit of
large circumference. The detector's response 
has no infrared ambiguity for $d=2$, 
neither in Minkowski nor on the cylinder. 
\end{abstract}

% \singlespacing

\newpage

\section{Introduction\label{sec:intro}}

In quantum field theory, the interaction between a scalar field and an 
observer is often studied by modelling the observer by 
a spatially pointlike system with discrete energy levels, 
an Unruh-DeWitt detector~\cite{Unruh:1976db,DeWitt:1979}. 
Despite its mathematical simplicity, 
this modelling captures the core features of the dipole 
interaction by which atomic orbitals couple to 
the electromagnetic field~\cite{MartinMartinez:2012th,Alhambra:2013uja}. 
In the special case of a uniformly linearly accelerated observer 
coupled to a field in its Minkowski vacuum, detector analyses 
have provided significant evidence that the Unruh effect~\cite{Unruh:1976db}, 
the thermal response of the observer, 
occurs whenever the 
interaction time is long, the interaction switch-on and switch-off are sufficiently slow 
and the back-reaction of the observer on the quantum field remains small 
\cite{Unruh:1976db,DeWitt:1979,Birrell:1982ix,Unruh:1983ms,Wald:1995yp,Higuchi:1993cya,Sriramkumar:1994pb,DeBievre:2006px,Lin:2006jw,Satz:2006kb,Crispino:2007eb,Louko:2007mu,Dappiaggi:2009fx,hodgkinson-louko,Barbado:2012fy,Juarez-Aubry:2014jba,Fewster:2016ewy}. 

In this paper we consider an Unruh-DeWitt detector coupled to a Dirac field, 
taking the interaction Hamiltonian to be 
linear in the Dirac field's scalar density, 
$\overline{\psi}\psi$
\cite{Iyer:1980yc,Takagi:1986kn,Langlois:2005nf,Langlois:2005if,Hummer:2015xaa}. 
The product of $\overline{\psi}$ and $\psi$ at the same spacetime 
point makes this interaction more singular than the 
conventional linear coupling to a scalar field~\cite{Unruh:1976db,DeWitt:1979}. 
Working in linear perturbation theory for a massive Dirac field, 
the detector's response has a divergent additive term, 
and although in stationary situations this term been viewed as 
a formally divergent constant that should be dropped 
in the dual limit of long interaction and small coupling~\cite{Takagi:1986kn}, 
in nonstationary situations the response 
would need an additional regularisation, 
perhaps by a spatial profile or by an appropriate normal ordering~\cite{Hummer:2015xaa,Suzuki:1997cz}. 
In the special case of Minkowski vacuum, the divergent 
term is however proportional to the mass of the field, 
and for a massless field a consistent regularisation is 
accomplished by simply dropping the 
additive term~\cite{Langlois:2005nf,Langlois:2005if}. 
In this paper we therefore focus on the massless field. 

Our first objective is to evaluate the detector's response 
on an arbitrary trajectory in Minkowski spacetime of dimension $d\ge2$ 
when the field is initially prepared in Minkowski vacuum, working in 
linear perturbation theory and allowing the detector to be switched on 
and off in an arbitrary smooth way. We show that the response is 
identical to that of a detector coupled linearly to a massless scalar 
field in $2d$ spacetime dimensions. In the special case of uniform 
linear acceleration, the long time limit of the detector's response 
hence exhibits the Unruh effect with a Planckian factor 
for all~$d$. By contrast, the Rindler power spectrum of the Dirac 
field is known to have a Planckian factor for odd $d$ but a 
Fermi-Dirac factor for even~$d$~\cite{Takagi:1986kn}. 
These observations are compatible since the detector's 
response is not equal to the Rindler power spectrum but is given by 
the convolution of the Rindler power spectrum with 
itself~\cite{Takagi:1986kn}. 

Our second objective is to consider a detector on an arbitrary 
worldline on a $(1+1)$-dimensional flat static cylinder. 
The main issue here is that the field has two spin structures, 
often referred to as the periodic field and the antiperiodic field, 
and while the antiperiodic field has a Minkowski-like Fock vacuum, 
the zero mode of the periodic field does not have a Fock vacuum. 
We evaluate the detector's response, showing that the response 
distinguishes the periodic and antiperiodic spin
structures, and the zero mode of the periodic spinor contributes to the 
response by a state-dependent but well defined way. 
We also give a selection of analytic and numerical results 
for inertial and uniformly accelerated
trajectories, recovering the $d=2$ Minkowski results in the limit of
large circumference. 

In two dimensions, our results show that the detector's 
response has no infrared ambiguity, 
neither in Minkowski nor on the cylinder. 
In this respect the massless Dirac field differs from the massless scalar field, whose 
response in two-dimensional Minkowski vacuum is ambiguous 
due to the additive ambiguity in the Wightman function~\cite{Decanini:2005eg}. 

We begin by recalling in Section \ref{sec:UDW-detector}
the definition of the field-detector model with 
an interaction Hamiltonian that is linear in the Dirac field's scalar density. 
The response in Minkowski vacuum in $d\ge2$ dimensions is evaluated in Section \ref{sec:Mink-vac}
and the response on the $(1+1)$-dimensional flat static cylinder in Section~\ref{sec:cylinder}. 
Inertial and uniformly accelerated trajectories on the cylinder are analysed in Section~\ref{sec:sample-trajectories}. Section \ref{sec:conclusions} gives a summary and brief concluding remarks. 
The spinorial conventions and notation are collected in Appendix~\ref{AppenGamma}, 
and a selection of technical calculations are deferred to 
Appendices \ref{AppenB}--\ref{app:stationary1+1}. 

We use units in which $\hbar =c =1$. The spacetime signature is mostly minus, 
$(+ - - \cdots)$. 
Spacetime points are denoted by math italic letters. In Minkowski spacetime, 
spacetime vectors are denoted by math italic letters and 
spatial vectors in a given Lorentz frame are denoted
by boldface letters. 
Overline on a scalar denotes the complex conjugate and 
overline on a spinor denotes the Dirac conjugate. 
% $\overline{\psi} = \psi^\dag \gamma^0$. 
$o(1)$~denotes a quantity that 
tends to zero in the limit under consideration.

\section{Unruh-DeWitt detector coupled to the Dirac field\label{sec:UDW-detector}}

In this section we briefly recall relevant properties 
of an Unruh-DeWitt detector that is 
coupled linearly to the scalar density of a Dirac field. 

We consider a pointlike detector that moves in a (possibly) 
curved spacetime on the worldline 
$x(\tau)$, where $\tau$ is the proper time. 
The detector is a two-level system, 
with the Hilbert space~$\BbbC^2$, 
spanned by the orthonormal basis 
$\{|E_0\rangle, |E_1\rangle\}$ of eigenstates of the Hamiltonian~$H_{D}$: 
$H_D |E_i\rangle = E_i |E_i\rangle$, where the 
eigenenergies $E_0$ and $E_1$ 
are real-valued constants. 

The detector is coupled to a Dirac field $\psi$ by the interaction Hamiltonian 
\begin{equation}
H_{\text{int}} :=  
c \chi(\tau) m(\tau) \overline{\psi} \big(x(\tau)\big) \psi\big(x(\tau)\big)
\ , 
\label{InterHamilt}
\end{equation}
where $m(\tau)$ is the detector's 
monopole moment operator, 
evolving in the interaction picture by 
\begin{align}
m(\tau) = e^{iH_{D}\tau}m(0)e^{-iH_{D}\tau}
\ , 
\end{align}
$c$~is a coupling constant, 
and the switching function $\chi$ is a smooth real-valued function 
that specifies how the interaction is turned on and off. 
We assume $\chi$ either to have compact support or 
to have so rapid falloff that the system can be treated as 
uncoupled in the asymptotic past and future. 

Before the interaction begins, 
the detector occupies the eigenstate 
$|E_{0}\rangle$ and the field occupies some Hadamard state~$|\Psi_{0}\rangle$. 
Working to linear order in~$c$, 
the probability for the detector to be found in the state 
$|E_{1}\rangle$ after the interaction has ceased, 
regardless the final state of the field, is 
\begin{align}
P(\Omega) = |c|^2 |\langle E_1|m(0)|E_0 \rangle|^2 \, 
\mathcal{F}(\Omega)
\ , 
\end{align}
where $\Omega = E_1-E_0$, 
the detector's response function $\mathcal{F}(\Omega)$ 
is given by 
\begin{align}
\mathcal{F}(\Omega) := 
\int 
d\tau \, d\tau' \, 
\chi(\tau) \chi(\tau') 
\, e^{-i\Omega(\tau-\tau')} \, W^{(2,\overline2)} \bigl(x(\tau), x(\tau') \bigr)
\ ,
\label{RespFunct-def}
\end{align}
and 
\begin{align}
W^{(2,\overline2)}(x, y) 
:= 
\langle \Psi_{0}| 
\overline{\psi}(x)
\psi(x)
\overline{\psi}(y)
\psi(y)
|\Psi_{0}\rangle
\ . 
\label{eq:W2-gen-def}
\end{align}
The factor $|c|^2 |\langle E_1 |m(0) |E_0\rangle|^2$ 
depends only on the inner working of the detector, and 
we drop it from now on, 
referring to the response function as the probability. 
Note that $\mathcal{F}(\Omega)$ gives the probability of an excitation for 
$\Omega > 0$ and the probability of a de-excitation for 
$\Omega < 0$. 

Although $|\Psi_{0}\rangle$ is by assumption Hadamard, 
formula \eqref{eq:W2-gen-def} as it stands does not define 
$W^{(2,\overline2)}\bigl(x(\tau), x(\tau^\prime)\bigr)$ as a 
distribution on the detector's worldline 
because of the partial coincidence limit in \eqref{eq:W2-gen-def} 
\cite{Takagi:1986kn,Langlois:2005nf,Langlois:2005if,Hummer:2015xaa}. 
To make the response function \eqref{RespFunct-def} well defined, 
it will be necessary to give formula \eqref{eq:W2-gen-def} 
an appropriate distributional interpretation. 
We shall address this in Sections \ref{sec:Mink-vac}
and \ref{sec:cylinder} below.

\section{Response in Minkowski vacuum\label{sec:Mink-vac}}

In this section we evaluate the detector's response 
to a massless Dirac field in Minkowski spacetime of dimension $d\ge2$, 
with the field in the usual Minkowski vacuum. 
We first recall relevant properties of the massive field, 
and we then show that the massless limit of the 
correlation function $W^{(2,\overline2)}(x,y)$ 
\eqref{eq:W2-gen-def} 
can be interpreted as a distribution 
% \cite{Takagi:1986kn,Langlois:2005nf,Langlois:2005if,Hummer:2015xaa} 
for which the response function \eqref{RespFunct-def}
is well defined.

\subsection{Quantum Dirac field}

We first recall some basic facts and notation about 
a massive Dirac field on Minkowski spacetime. 

We denote the spacetime points by $x=(x^0,\textbf{x}) =(t,\textbf{x})$,
and the Minkowski metric is 
$\eta_{\mu\nu} = \diag(1,-1,-1,\ldots)$. 
The action of the Dirac field $\psi$ is 
\begin{align}
S = \int d^{d}x \, 
\overline{\psi}(x) \left(i\gamma^\mu \partial_\mu -m \right)\psi(x)
\ , 
\end{align} 
where $m>0$ is the mass and the conventions for the gamma matrices 
$\gamma^\mu$,  $\mu=0,1,2,\ldots,d-1$, 
are summarised in Appendix~\ref{AppenGamma}. 
The field equations 
for $\psi$ and its 
Dirac conjugate $\overline{\psi} = \psi^\dagger \gamma^0$ are 
\begin{subequations}
\begin{align}
0 &= (i\gamma^\mu\partial_\mu - m)\psi(x)
\ , 
\label{eq:Dirac-equation}
\\
0 &= 
i\partial_\mu\overline{\psi}(x)\gamma^\mu + m\overline{\psi}(x)
\ . 
\end{align}
\end{subequations}

A complete set of mode solutions to \eqref{eq:Dirac-equation} is 
\begin{subequations}
\label{IndepeSolDirac}
\begin{align}
\label{eq1op}  
u_{\textbf{k}}^{(s)}(x) & = u^{(s)}(\textbf{k}) e^{-ikx}
\ ,
\\
\label{eq2op}  
v_{\textbf{k}}^{(s)}(x) & = v^{(s)}(\textbf{k})e^{ikx}
\ ,
\end{align}
\end{subequations}
where 
$k = (k^0, \textbf{k})$, 
$k^0 := \omega_{\textbf{k}} = \bigl(\textbf{k}^2+m^2\bigr)^{1/2}$, 
$kx = k^0 x^0 - \textbf{k} \cdot \textbf{x}$, 
and the spinors $u^{(s)}(\textbf{k})$ and $v^{(s)}(\textbf{k})$ 
are as given in Appendix~\ref{AppenGamma}, 
with $s$ being the helicity index. 
In the Dirac inner product, given by 
\begin{align}
\langle\psi,\phi \rangle = 
\int d^{d-1}x
\, \overline{\psi}(t,\textbf{x})\gamma_0\phi(t,\textbf{x})
\ , 
\label{eq:Dirac-inner}
\end{align} 
these mode solutions are normalised to 
\begin{subequations}
\begin{align}
& 
\bigl\langle u_{\textbf{k}}^{(s)} , u_{\textbf{k}'}^{(s')} \bigr\rangle
= 
\bigl\langle v_{\textbf{k}}^{(s)} , v_{\textbf{k}'}^{(s')} \bigr\rangle
= 
2\omega_{\textbf{k}} (2\pi)^{d-1} 
\delta^{s s'} \delta^{d-1}(\textbf{k}-\textbf{k}^\prime)
\ , 
\\
& 
\bigl\langle u_{\textbf{k}}^{(s)} , v_{\textbf{k}'}^{(s')} \bigr\rangle=0
\ . 
\end{align}
\end{subequations}

The quantised field is expanded as 
\begin{align} 
\label{eq:psi-quantum-expansion}
\psi(x)
=\int \widetilde{dk}\sum_{s}
\Bigl(
b_s(\textbf{k}) u_{\textbf{k}}^{(s)}(x) 
+ d_s^\dagger(\textbf{k}) v_{\textbf{k}}^{(s)}(x)
\Bigr)
\ , 
\end{align}
where 
\begin{align} 
\widetilde{dk} := \frac{d^{d-1}k}{2\omega_{\textbf{k}} (2\pi)^{d-1}}
\ ,
\end{align}
and the only nonvanishing anticommutators of the operator coefficients are
\begin{align}
\bigl\lbrace b_s(\textbf{k}),b_{s'}^\dagger(\textbf{k}^\prime)\bigr\rbrace 
=\bigl\lbrace d_s(\textbf{k}),d^\dagger_{s'}(\textbf{k}^\prime)\bigr\rbrace
= 2\omega_{\textbf{k}} (2 \pi)^{d-1} \delta_{s s'} 
\, \delta^{d-1}(\textbf{k}-\textbf{k}^\prime)
\ . 
\end{align}
The field's equal-time anticommutators are 
\begin{subequations}
\begin{align} 
& \bigl\{\psi_a(t,\mathbf{x}),\psi_b(t,\mathbf{x}') \bigr\} 
= 
\bigl\{\psi^{\dagger}_a(t,\mathbf{x}),\psi^{\dagger}_b(t,\mathbf{x}') \bigr\} 
= 0 
\ , 
\\
& \bigl\{\psi_a(t,\mathbf{x}),\psi^{\dagger}_b(t,\mathbf{x}') \bigr\} 
= 
\delta_{ab} \delta^{d-1}(\mathbf{x} - \mathbf{x}')
\ , 
\end{align}
\end{subequations}
where we have explicitly written out the spinor indices. 
The fermionic Fock space is built on the Minkowski 
vacuum state $|0\rangle$ which satisfies 
$b_s(\textbf{k}) |0\rangle = d_s(\textbf{k}) |0\rangle = 0$. 

$\psi$ and $\overline{\psi}$ may be decomposed into their 
positive and negative frequency components as 
\begin{subequations}
\begin{align}
\psi(x) &= \psi^+(x) + \psi^-(x)
\ ,
\\
\overline{\psi}(x) &= 
% \psi^\dagger\gamma^0 = 
\overline{\psi}^+(x) + \overline{\psi}^-(x)
\ ,
\end{align}
\end{subequations}
where
\begin{subequations}
\label{eq:psi-andbar-pm}
\begin{align}
\psi^+(x)&=\int \widetilde{dk} \sum_s b_s(\textbf{k})
u_{\textbf{k}}^{(s)}(x) 
\ , 
\\ 
\psi^-(x) &=
\int \widetilde{dk} \sum_s
d_s^\dagger(\textbf{k})v_{\textbf{k}}^{(s)}(x)
\ , 
\\
\overline{\psi}^+(x) &= 
\int \widetilde{dk} \sum_s d_s(\textbf{k})\overline{v}_{\textbf{k}}^{(s)}(x)
\ , 
\\
\overline{\psi}^-(x) &= 
\int \widetilde{dk}\sum_s b_s^\dagger(\textbf{k}) \overline{u}_{\textbf{k}}^{(s)}(x)
\ . 
\end{align}
\end{subequations}
In the conventions of~\cite{Takagi:1986kn}, 
the Dirac field Wightman functions $S^\pm(x, y)$ are 
\begin{subequations}
\label{eq:bothWigFunct}
\begin{align}
S_{ab}^+(x, y) &:= 
\langle 0 | \psi_a (x) \overline{\psi}_b(y) |0\rangle 
= 
\bigl\lbrace\psi_a^+(x), \overline{\psi}_b^-(y)\bigr\rbrace  
\notag
\\
& 
\,\, = \left(i\gamma^\mu \partial_{x^\mu} + m\right)_{ab} G^+(x, y)
\ , 
\label{PosWigFunct}
\\
S_{ab}^-(x, y) & := 
\langle 0 | \overline{\psi}_b(y) \psi_a(x) |0\rangle 
= 
\bigl\lbrace \psi_a^-(x),\overline{\psi}_b^+(y) \bigr\rbrace
\notag
\\
& 
\,\, = - \left(i\gamma^\mu \partial_{x^\mu} + m\right)_{ab} G^+(y, x)
\ , 
\label{negWigFunct}
\end{align}
\end{subequations}
where $G^{+}$ is the Wightman function of a real scalar field of mass~$m$, 
\begin{align}
G^+(x, y)
= \int \widetilde{dk} \, e^{-ik(x - y)} 
\ , 
\label{eq:iDeltapm-integral}
\end{align}
and the distributional sense in \eqref{eq:iDeltapm-integral} is that of 
$x^0-y^0 \to x^0-y^0 - i \epsilon$ and the limit $\epsilon \to 0_+$. 
The explicit expression for $G^+(x, y)$ is \cite{Takagi:1986kn} 
\begin{align}
G^+(x, y)
= 
\frac{1}{2\pi}\left(\frac{m}{2\pi z(x,y)} \right)^{d/2-1}K_{d/2 -1}\bigl(mz(x,y)\bigr)
\ , 
\label{eq:scalar-wightman-massive}
\end{align}
where $K_{d/2-1}$ is the modified Bessel function of the second kind \cite{dlmf} and  
\begin{align}
z(x,y) := \sqrt{ (\textbf{x}-\textbf{y})^2 -(x^0-y^0-i\epsilon)^2}
\ . 
\label{eq:z-diff-def}
\end{align}

\subsection{$W^{(2,\overline2)}(x,y)$\label{subsec:Mink-W2}}

We wish to examine the 
correlation function $W^{(2,\overline2)}(x,y)$~\eqref{eq:W2-gen-def}. 

We show in Appendix \ref{AppenB} that 
\begin{align}
W^{(2,\overline2)} (x, y) 
= 
\Tr \bigl[ S^+(x,y) S^-(y,x) \bigr]
+ \Tr \bigl[ S^-(0,0) \bigr] 
\Tr \bigl[ S^-(0,0) \bigr] 
\ . 
\label{eq:CorrFunct-final-fromapp}
\end{align} 
Using $\Tr(\gamma^\mu)=0$ and $\Tr(\gamma^\mu \gamma^\nu)= N_d \eta^{\mu\nu}$, 
\eqref{eq:bothWigFunct}
and \eqref{eq:scalar-wightman-massive}
show that the first term in 
\eqref{eq:CorrFunct-final-fromapp} can be written as 
\begin{align}
% \label{eq:CorrFunct-final-fromapp}
\Tr \bigl[ S^+(x,y) S^-(y,x) \bigr]
& = 
\frac{N_d \, m^2}{(2\pi)^2} \left(\frac{m}{2\pi z(x,y)}\right)^{d-2} 
\times 
\notag
\\
& \hspace{4ex}
\times 
\left\lbrace
\left[K_{d/2}\bigl(mz(x,y)\bigr)\right]^2 
- 
\left[K_{d/2 -1}\bigl(mz(x,y)\bigr)\right]^2 
\right\rbrace
\ , 
\label{eq:W2-firstterm-massive}
\end{align}
which is a well-defined distribution. 
In the second term in~\eqref{eq:CorrFunct-final-fromapp}, by contrast, we have, 
using \eqref{negWigFunct} and $\Tr(\gamma^\mu)=0$, 
\begin{align}
\Tr \bigl[ S^-(x,y) \bigr] = - N_d m G^{+}(y, x)
\ , 
\label{eq:TrSminus}
\end{align}
which diverges as $(x,y) \to (0,0)$ by~\eqref{eq:scalar-wightman-massive}. 
$W^{(2,\overline2)} (x, y)$ is hence not well defined, 
due to a divergent additive constant in the 
second term in \eqref{eq:CorrFunct-final-fromapp}
\cite{Takagi:1986kn,Langlois:2005nf,Langlois:2005if,Hummer:2015xaa}. 

Consider however now the limit $m\to0$. 
If the second term in \eqref{eq:CorrFunct-final-fromapp} 
is dropped in this limit, 
we obtain 
\begin{align}
W^{(2,\overline2)} (x, y) 
& = 
\frac{N_d \bigl(\Gamma(d/2)\bigr)^2}{4\pi^{d} \bigl(z(x,y)\bigr)^{2d-2}}
\ , 
\label{eq:CorrFunct-final-massless}
\end{align}
using \eqref{eq:W2-firstterm-massive} and the small argument form of the 
modified Bessel function~\cite{dlmf}. 
We adopt \eqref{eq:CorrFunct-final-massless} as the definition 
of $W^{(2,\overline2)}$ for the massless field. 

We shall not attempt to justify dropping the 
second term in \eqref{eq:CorrFunct-final-fromapp} as $m\to0$ 
from some underlying framework that would provide a definition for the 
coincidence limit of a squared distribution, 
but we can make two consistency observations. 

First, from \eqref{eq:scalar-wightman-massive}, 
\eqref{eq:TrSminus} 
and the small argument form of the 
modified Bessel function \cite{dlmf} 
we see that $\Tr \bigl[ S^-(x,y) \bigr]$ 
has a well defined distributional limit as $m\to0$, 
and this limit is the zero distribution. 

Second, recall that the Wightman function of a massless scalar field is given by 
\cite{Decanini:2005eg} 
\begin{align}
\label{eq:scalar-wightman-massless}
G^+(x, y)
= 
\begin{cases}
\displaystyle{
\frac{\Gamma\bigl(\frac{d}{2}-1\bigr) \vphantom{A^{\displaystyle{A}}}}{4\pi^{d/2} \bigl(z(x,y)\bigr)^{d-2}}}
&
\text{for $d>2$}, 
\\[3ex]
\displaystyle{
- (2\pi)^{-1} \ln \bigl(\mu z(x,y)\bigr)}
&
\text{for $d=2$}, 
\end{cases}
\end{align}
where $\mu$ is an undetermined positive constant of dimension inverse length. 
For $d>2$, \eqref{eq:scalar-wightman-massless} is obtained as the $m\to0$ 
limit of~\eqref{eq:scalar-wightman-massive}. 
For $d=2$, 
\eqref{eq:scalar-wightman-massless} is obtained as the $m\to0$ 
limit of \eqref{eq:scalar-wightman-massive} after subtracting an 
$m$-dependent constant that diverges as $m\to0$, 
and the arbitrariness (``infrared ambiguity'')
in this subtraction is encoded 
in the positive constant $\mu$ in~\eqref{eq:scalar-wightman-massless}. 
Substituting \eqref{eq:scalar-wightman-massless} in 
\eqref{eq:bothWigFunct} with $m=0$ gives $S^\pm(x,y)$ 
such that $\Tr \bigl[ S^-(x,y) \bigr]$ vanishes as a distribution, 
and substituting these $S^\pm(x,y)$ in the first term in 
\eqref{eq:CorrFunct-final-fromapp} and $\Tr \bigl[ S^-(0,0) \bigr]$ in the second term in 
\eqref{eq:CorrFunct-final-fromapp} 
gives~\eqref{eq:CorrFunct-final-massless}. 

Note that $W^{(2,\overline2)} (x,y)$ \eqref{eq:CorrFunct-final-massless}
tends to zero in the limit of large spacelike separation for all $d\ge2$, including $d=2$. 
For $d=2$, $W^{(2,\overline2)}$ has no infrared ambiguity, 
in contrast to the infrared ambiguity of 
$G^+$~\eqref{eq:scalar-wightman-massless}.

\subsection{Detector's response to a massless field}

Collecting \eqref{RespFunct-def} and~\eqref{eq:CorrFunct-final-massless}, 
we see that the detector's response to a massless Dirac field is given by 
\begin{align}
\mathcal{F}(\Omega) = 
\frac{N_d \bigl(\Gamma(d/2)\bigr)^2}{4\pi^{d}}
\int
d\tau \, d\tau' 
\, 
\frac{\chi(\tau) \chi(\tau') 
\, e^{-i\Omega(\tau-\tau^\prime)}}{\big[z\bigl(x(\tau), x(\tau')\bigr)\bigr]^{2d-2}}
\ ,
\label{Mink-RespFunct-massless-final}
\end{align}
where we recall from \eqref{eq:z-diff-def} that 
\begin{align}
z(x,y) = \sqrt{ (\textbf{x}-\textbf{y})^2 -(x^0-y^0-i\epsilon)^2}
\label{eq:z-diff-recap}
\end{align}
with $\epsilon\to0_+$. This result agrees with the limits, 
special cases and alternative forms considered in 
\cite{Takagi:1986kn,Langlois:2005nf,Langlois:2005if,Hummer:2015xaa}. 

To set this result in context, recall that 
the response of an Unruh-DeWitt detector 
that is linearly coupled to a scalar field in its Minkowski vacuum 
is \cite{Birrell:1982ix,Wald:1995yp,Junker} 
\begin{align}
\label{RespFunct-scalar-gen}
\mathcal{F}_{\text{sc}}(\Omega) = 
\int d\tau \, d\tau' \, 
\chi(\tau) \chi(\tau') 
\, e^{-i\Omega(\tau-\tau')} \, G^+ \bigl(x(\tau), x(\tau') \bigr)
\ , 
\end{align}
where $G^+$ is the scalar field's Wightman function. 
By \eqref{eq:scalar-wightman-massless}, 
\eqref{Mink-RespFunct-massless-final}
and 
\eqref{RespFunct-scalar-gen}, we may hence formalise our observations 
as the following theorem: 

\begin{thm}
The response function of an Unruh-DeWitt detector 
coupled quadratically 
to a massless Dirac field in Minkowski vacuum in $d\ge2$ 
spacetime dimensions equals 
\begin{align}
\frac{N_d \bigl(\Gamma(d/2)\bigr)^2}{\Gamma(d-1)}
\end{align}
times the response function of an Unruh-DeWitt detector
coupled linearly to a massless scalar field in Minkowski 
vacuum in $2d$ spacetime dimensions. 
\label{thm:massless-fermion-to-scalar}
\end{thm}

One consequence of Theorem \ref{thm:massless-fermion-to-scalar}
is that the Dirac field detector's response is 
well defined whenever the detector's 
worldline is smooth, by the corresponding 
result for the scalar field detector~\cite{Hormander:1983,Fewster:1999gj}. 

The special case of a uniformly linearly accelerated detector deserves a comment. 
In the limit in which the detector operates for a long time and the 
switching effects are negligible, 
it is well known \cite{Takagi:1986kn} 
that both the response function of the 
scalar field detector and the response function of the Dirac field detector 
satisfy the detailed balance condition, 
\begin{align}
\mathcal{F}(-\Omega) = e^{\Omega/T_{\text{U}}} \mathcal{F}(\Omega) 
\ , 
\end{align}
where $T_{\text{U}} := a/(2\pi)$ and $a$ is the magnitude of the 
detector's proper acceleration. 
This is the celebrated Unruh effect, and $T_{\text{U}}$ 
is the Unruh temperature~\cite{Unruh:1976db}. 
It was observed in \cite{Takagi:1986kn} 
that the response function 
of the scalar field detector involves 
a Planck factor in even spacetime dimensions 
but a Fermi-Dirac factor in odd spacetime dimensions. 
Theorem \ref{thm:massless-fermion-to-scalar} 
hence implies that the response function 
of the Dirac field detector involves 
a Planck factor in \emph{all\/} spacetime dimensions.

By contrast, recall that the ``Rindler noise'' of the Dirac field, 
defined as a Fourier transform of the Wightman function 
$S^+$ over the uniformly accelerated trajectory, 
involves a Fermi-Dirac factor in even spacetime dimensions and 
a Planck factor in odd spacetime dimensions~\cite{Takagi:1986kn}. 
This is fully compatible with our observation that the detector's 
response involves a Planck factor in all spacetime dimensions: 
the response function is not directly the Rindler noise but rather the 
self-convolution of the Rindler noise, as shown in 
(8.5.13) in~\cite{Takagi:1986kn}, 
and a Fermi-Dirac factor 
in the Rindler noise does not imply a Fermi-Dirac 
factor in the response function. 
We have explicitly checked that 
our Theorem \ref{thm:massless-fermion-to-scalar} agrees with 
(8.5.13) in \cite{Takagi:1986kn} 
for a Dirac field  
in spacetime dimensions 2, 3 and~4. 
The verbal description of the 
Fermi-Dirac versus Planck factors 
in the Dirac field detector's response function
given in \cite{Takagi:1986kn}, 
in the full paragraph between
(8.5.14) and (8.5.15), is hence not accurate.

\section{Cylindrical $(1+1)$-dimensional spacetime\label{sec:cylinder}}

In this section we consider a detector coupled to a massless Dirac field
in a flat static cylindrical spacetime in $1+1$ dimensions. 
The main new issue is that there are now two inequivalent spin structures, 
and one of the spin structures has a zero mode.

\subsection{Massive Dirac field on the cylindrical spacetime}

The spacetime is a flat static $(1+1)$-dimensional 
cylinder with spatial circumference $L>0$. 
We work in standard Minkowski coordinates 
$(t,x)$ in which the metric reads 
\begin{align}
ds^2 = dt^2 - dx^2
\ , 
\label{eq:1+1-minkmetric}
\end{align}
with the periodic identification $(t,x) \sim (t,x+L)$. 

We consider a Dirac field $\psi$ with mass $m>0$. 
We use the Minkowski spacetime notation 
of Section \ref{sec:Mink-vac} 
with the exception that $\psi$ is now 
either periodic or antiperiodic as $(t,x) \mapsto (t,x+L)$. 
The choice of periodicity versus antiperiodicity 
implements the choice between the two inequivalent 
spin structures of the field~\cite{Birrell:1982ix}: 
we refer to these spin structures as respectively the 
\emph{periodic\/} or \emph{untwisted\/} 
spin structure and the 
\emph{antiperiodic\/} or \emph{twisted\/} 
spin structure. 
We suppress explicit references to the spin 
structure in the formulas until the final 
expressions for the detector's response in 
\eqref{GeneralRFTS} and~\eqref{eq:Funwisted-general}. 

A complete set of mode solutions for each spin structure is 
\begin{subequations}
\label{eq:2dim-modesols}
\begin{align}
u_n(t,x) &= (2L \omega_n)^{-1/2} \, u(k_n) \, e^{-i(\omega_n t - k_n x)}
\ , 
\\
v_n(t,x) &= (2L \omega_n)^{-1/2} \, v(k_n) \, e^{i(\omega_n t - k_n x)}
\ , 
\end{align}
\end{subequations}
where $n\in\BbbZ$ and 
\begin{subequations}
\label{eq:cylindermomenta}
\begin{align}
k_n &:= \begin{cases}
2\pi n /L & \text{for untwisted spinors}, 
\\
2\pi(n+\tfrac12)/L & \text{for twisted spinors}, 
\end{cases}
\\[1ex]
\omega_n & := \bigl( m^2 + k_n^2 \bigr)^{1/2}
\ , 
\end{align}
\end{subequations}
and the spinors $u(k_n)$ and $v(k_n)$ are the $(1+1)$-dimensional 
special case of the spinors introduced in Appendix~\ref{AppenGamma}. 
Note that the spinors carry no spin index. 
The Dirac inner product \eqref{eq:Dirac-inner} 
is modified to 
\begin{align}
\langle\psi,\phi \rangle = 
\int_0^L dx
\, \overline{\psi}(t,\textbf{x})\gamma_0\phi(t,\textbf{x})
\ , 
\end{align} 
in which the mode solutions \eqref{eq:2dim-modesols} are normalised to 
\begin{subequations}
\begin{align}
& 
\bigl\langle u_n , u_{n'} \bigr\rangle
= 
\bigl\langle v_n , v_{n'} \bigr\rangle
= \delta_{n n'}
\ , 
\\
& 
\bigl\langle u_n , v_{n'} \bigr\rangle=0
\ . 
\end{align}
\end{subequations}

The quantised field is expanded as 
\begin{equation}
\psi(t,x) = \sum_n
\bigl( 
b_n u_n(t,x) +d_n^{\dagger} v_n(t,x) 
\bigr)
\ ,
\label{eq:psi-expansion-cylinder}
\end{equation}
where the only nonvanishing anticommutators of the coefficients are 
\begin{align}
\bigl\lbrace b_n, b_{n'}^\dagger \bigr\rbrace 
=\bigl\lbrace d_n ,d^\dagger_{n'} \bigr\rbrace
= \delta_{n n'} 
\ . 
\label{eq:cylinder-b-d-anticomm}
\end{align}
The field's equal-time anticommutators are 
\begin{subequations}
\label{eq:cylinder-et-anticomm}
\begin{align} 
& \bigl\{\psi_a(t,x),\psi_b(t,x') \bigr\} 
= 
\bigl\{\psi^{\dagger}_a(t,x),\psi^{\dagger}_b(t,x') \bigr\} 
= 0 
\ , 
\\
& \bigl\{\psi_a(t,x),\psi^{\dagger}_b(t,x') \bigr\} 
= 
\delta_{ab} \delta(x,x')
\ , 
\end{align}
\end{subequations}
where $\delta(x,x')$ denotes Dirac's delta-function on the circle. 
The fermionic Fock space is built on the 
vacuum state $|0\rangle$ which satisfies 
$b_n |0\rangle = d_n |0\rangle = 0$. 

Proceeding as in Section~\ref{sec:Mink-vac}, we have 
\begin{subequations}
\label{eq:bothWigFunct-2dim}
\begin{align}
S_{ab}^+(t,x; t',x') &:= 
\langle 0 | \psi_a (t,x) \overline{\psi}_b(t', x') |0\rangle 
= \left(i\gamma^\mu \partial_{\mu} + m\right)_{ab} G^+(t,x; t',x')
\ , 
\label{PosWigFunct-2dim}
\\
S_{ab}^-(t,x; t',x') & := 
\langle 0 | \overline{\psi}_b(t', x') \psi_a(t,x) |0\rangle 
= - \left(i\gamma^\mu \partial_{\mu} + m\right)_{ab} G^+(t',x'; t,x)
\ , 
\label{negWigFunct-2dim}
\end{align}
\end{subequations}
where 
\begin{align}
G^+(t,x ; t', x')
= \sum_n 
\frac{1}{2\omega_n L} 
\exp \bigl[ -i\omega_n(t-t' - i\epsilon) + i k_n (x-x') \bigr] 
\ , 
\label{eq:Gplus-cylinder-sum}
\end{align}
understood in the sense $\epsilon\to0_+$, 
and the differentiation in \eqref{eq:bothWigFunct-2dim} is with respect to the unprimed argument. 
For the untwisted spinor, $G^{+}$ is the Wightman function of a real scalar field of mass~$m$. 
For the twisted spinor, $G^{+}$ is the Wightman function of a
scalar field that takes values on a twisted $\BbbR/\BbbZ_2$ bundle~\cite{Birrell:1982ix}. 

The correlation function $W^{(2,\overline2)}$ \eqref{eq:W2-gen-def} is again given by 
the ill-defined expression~\eqref{eq:CorrFunct-final-fromapp}. 
We shall give a well-defined interpretation for this expression in the massless limit 
for each of the two spin structures in turn.

\subsection{Twisted massless field}

Consider the twisted massless Dirac field. 
Proceeding as in subsection~\ref{subsec:Mink-W2}, we drop the second 
term in \eqref{eq:CorrFunct-final-fromapp} 
and use in the first term \eqref{eq:bothWigFunct-2dim} with $m=0$, obtaining 
\begin{align}
W^{(2,\overline2)}_t (t,x; t', x') 
& =
\Tr \! \left\{ \bigl[S^+(t,x; t', x')\bigr]^2 \right\}
\notag 
\\
& =
-2 
\bigl[ (\partial_t - \partial_x) G^+(t,x; t', x') \bigr] 
\bigl[ (\partial_t + \partial_x) G^+(t,x; t', x') \bigr] 
\notag 
\\
& =
\frac{2}{L^2} e^{-2 \pi i (\Delta t-i\epsilon)/L}
\sum_{n,m=0}^{\infty}
e^{-2 \pi i n (\Delta t+ \Delta x -i\epsilon)/L} 
\, 
e^{-2 \pi i m (\Delta t- \Delta x -i\epsilon)/L}
\notag
\\
& =
- \frac{1}{2L^2 \sin\bigl[ \pi(\Delta t+ \Delta x -i\epsilon)/L \bigr]
\sin\bigl[ \pi(\Delta t- \Delta x -i\epsilon)/L \bigr]}
\ , 
\label{General2-pointCFTS}
\end{align}
where $\Delta t := t-t'$ and $\Delta x := x-x'$, and the subscript in $W^{(2,\overline2)}_t$ 
refers to the twisted spin structure. 

The response of the detector is obtained from \eqref{RespFunct-def} 
with~\eqref{General2-pointCFTS}. 
Note that the response contains no infrared ambiguities. 
A formula that is well suited for numerical evaluation 
is obtained by using the sum form in~\eqref{General2-pointCFTS}, 
yielding 
\begin{align}
\mathcal{F}_t(\Omega) 
&= 
\frac{2}{L^2}  
\sum_{n,m=0}^{\infty}
\int d\tau \, d \tau'  \, 
\chi(\tau)\chi(\tau')
\, e^{-i\Omega(\tau-\tau^\prime)}
\notag
\\[1ex]
&\hspace{3ex}
\times 
\exp \! \left(
-\frac{2 \pi i \left\{ (n+m+1)[t(\tau)-t(\tau^\prime) - i\epsilon]
+ (n-m)[(x(\tau)-x(\tau^\prime)] \right\}}{L}
\right) 
\ . 
\label{GeneralRFTS}
\end{align}

\subsection{Untwisted massless field}

Consider the untwisted massless Dirac field. 
As the $n=0$ term in 
\eqref{eq:psi-expansion-cylinder} has vanishing frequency, 
this mode does not have a Fock vacuum. 
We hence split $\psi$ as 
\begin{align}
\psi(t,x) &= \psi^{\text{osc}}(t,x) + \psi^{\text{zm}}(t) 
\ , 
\label{eq:psi-osc+zm-split}
\end{align}
where 
\begin{equation}
\psi^{\text{osc}}(t,x) = \sum_{n\ne0}
\bigl( 
b_n u_n(t,x) +d_n^{\dagger} v_n(t,x) 
\bigr)
\ ,
\label{eq:psi-osc-def}
\end{equation}
and $\psi^{\text{zm}}(t)$ is spatially constant. 
We treat $\psi^{\text{osc}}$ and $\psi^{\text{zm}}$ in turn and then combine the two. 

\subsubsection{Oscillator modes\label{subsubsec:oscillatormodes}}

We quantise the oscillator modes $\psi^{\text{osc}}$ with the usual 
anticommutators~\eqref{eq:cylinder-b-d-anticomm}. 
It follows that the equal-time anticommutators of 
$\psi^{\text{osc}}$ are 
\begin{subequations}
\label{eq:cylinder-et-osc-anticomm}
\begin{align} 
& \bigl\{\psi^{\text{osc}}_a(t,x),\psi^{\text{osc}}_b(t,x') \bigr\} 
= 
\bigl\{\psi^{\text{osc}\,\dagger}_a(t,x),\psi^{\text{osc}\,\dagger}_b(t,x') \bigr\} 
= 0 
\ , 
\\
& \bigl\{\psi^{\text{osc}}_a(t,x),\psi^{\text{osc}\,\dagger}_b(t,x') \bigr\} 
= 
\delta_{ab} \delta(x,x') - \delta_{ab}/L
\ . 
\label{eq:cylinder-et-osc-anticomm-psi-psidag}
\end{align}
\end{subequations}

Let $|0^{\text{osc}}\rangle$ denote the oscillator mode Fock vacuum, 
satisfying 
$b_n|0^{\text{osc}}\rangle = d_n|0^{\text{osc}}\rangle = 0$ for $n\ne0$. 
Proceeding as in~\eqref{eq:bothWigFunct-2dim}, we find 
\begin{subequations}
\label{eq:bothWigFunct-2-osc}
\begin{align}
S_{ab}^{\text{osc}+}(t,x; t',x') &:= 
\langle 0^{\text{osc}} | \psi_a^{\text{osc}} (t,x) 
\overline{\psi^{\text{osc}}_b}(t', x') |0^{\text{osc}}\rangle 
= i \left(\gamma^\mu \right)_{ab} \partial_\mu G^{\text{osc}+}(t,x; t',x')
\ , 
\label{eq:PosWigFunct-2-osc}
\\
S_{ab}^{\text{osc}-}(t,x; t',x') & := 
\langle 0^{\text{osc}} | \overline{\psi^{\text{osc}}_b}(t', x') 
\psi^{\text{osc}}_a(t,x) |0^{\text{osc}}\rangle 
= - i \left(\gamma^\mu \right)_{ab} \partial_{\mu} G^{\text{osc}+}(t',x'; t,x)
\ , 
\label{eq:NegWigFunct-2-osc}
\end{align}
\end{subequations}
where the differentiation is with respect to the unprimed argument and 
\begin{align}
G^{\text{osc}+}(t,x ; t', x')
= \sum_{n\ne0}
\frac{1}{2\omega_n L} 
\exp \bigl[ -i\omega_n(t-t' - i\epsilon) + i k_n (x-x') \bigr] 
\ . 
\label{eq:Gplus-cylinder-osc-sum}
\end{align}
Hence 
\begin{align}
W^{(2,\overline2)}_{\text{osc}}(t,x; t', x') 
& :=
\Tr \! \left[ S^{\text{osc}+}(t,x; t', x') S^{\text{osc}-}(t',x'; t, x)\right]
\notag 
\\
& =
\Tr \! \left\{ \bigl[S^{\text{osc}+}(t,x; t', x')\bigr]^2 \right\}
\notag 
\\
& =
-2 
\bigl[ (\partial_t - \partial_x) G^{\text{osc}+}(t,x; t', x') \bigr] 
\bigl[ (\partial_t + \partial_x) G^{\text{osc}+}(t,x; t', x') \bigr] 
\notag 
\\
& =
\frac{2}{L^2} 
\sum_{n,m=1}^{\infty}
e^{-2 \pi i n (\Delta t+ \Delta x -i\epsilon)/L} 
\, 
e^{-2 \pi i m (\Delta t- \Delta x -i\epsilon)/L}
\notag
\\
& = 
- \frac{\exp \bigl[-2 \pi i(\Delta t -i\epsilon)/L \bigr]}
{2L^2 \sin\bigl[ \pi(\Delta t+ \Delta x -i\epsilon)/L \bigr]
\sin\bigl[ \pi(\Delta t- \Delta x -i\epsilon)/L \bigr]}
\ . 
\label{eq:W2-osc-final}
\end{align}

\subsubsection{Zero mode\label{subsubsec:zeromode}}

We quantise the zero mode $\psi^{\text{zm}}$ so that 
$\psi^{\text{zm}}(t)$ and $\psi^{\text{zm}\,\dagger}(t)$ anticommute with 
$\psi^{\text{osc}}(t,x)$ and $\psi^{\text{osc}\,\dagger}(t,x')$ and satisfy 
\begin{subequations}
\label{eq:cylinder-et-zm-anticomm}
\begin{align} 
& \bigl\{\psi^{\text{zm}}_a(t),\psi^{\text{zm}}_b(t) \bigr\} 
= 
\bigl\{\psi^{\text{zm}\,\dagger}_a(t),\psi^{\text{zm}\,\dagger}_b(t) \bigr\} 
= 0 
\ , 
\\
& \bigl\{\psi^{\text{zm}}_a(t),\psi^{\text{zm}\,\dagger}_b(t) \bigr\} 
= 
\delta_{ab}/L
\ . 
\end{align}
\end{subequations}
Together with \eqref{eq:cylinder-et-osc-anticomm}, 
this ensures that the full Dirac field \eqref{eq:psi-osc+zm-split} 
satisfies the equal-time 
anticommutators~\eqref{eq:cylinder-et-anticomm}. 

Inserting \eqref{eq:psi-osc+zm-split} 
in the action shows that $\psi^{\text{zm}}$ is independent of~$t$. 
To satisfy~\eqref{eq:cylinder-et-zm-anticomm}, 
we write (cf.\ Chapter 20 of~\cite{henneaux-teitelboim-book})  
\begin{align}
\psi^{\text{zm}} 
= \frac{1}{\sqrt L}
\begin{pmatrix}Q_1\\Q^\dagger_2
\end{pmatrix}
\ ,
\end{align}
where $Q_a$ are independent of $t$ and satisfy 
\begin{subequations}
\label{eq:cylinder-et-zm-Q-anticomm}
\begin{align} 
& \bigl\{Q_a,Q_b \bigr\} = \bigl\{Q^\dagger_a,Q^\dagger_b \bigr\}
= 0 
\ , 
\\
& \bigl\{Q_a,Q^\dagger_b \bigr\} = 
\delta_{ab}
\ . 
\end{align}
\end{subequations}
The Hilbert space is built on the normalised state 
$|0^{\text{zm}}\rangle$ that satisfies $Q_a|0^{\text{zm}}\rangle=0$. 
The Hilbert space has dimension four, and an orthonormal basis is
$\bigl\{ |0^{\text{zm}}\rangle, 
Q^\dagger_1|0^{\text{zm}}\rangle, 
Q^\dagger_2|0^{\text{zm}}\rangle,
Q^\dagger_1 Q^\dagger_2|0^{\text{zm}}\rangle
\bigr\}$. 

For concreteness, we may work in a representation in which  
$\gamma^0 = \bigl( \begin{smallmatrix}0&1\\1&0\end{smallmatrix} \bigr)$ and 
$\gamma^1 = \bigl( \begin{smallmatrix}0&1\\-1&0\end{smallmatrix} \bigr)$. 
We then have 
\begin{align}
\overline{\psi^{\text{zm}}} \, \psi^{\text{zm}} 
= L^{-1} \bigl(Q_2 Q_1 +  Q^\dagger_1 Q^\dagger_2 \bigr)
\ . 
\end{align}
If the zero mode is in the normalised state 
\begin{align}
|\text{ZM}\rangle := 
\frac{\bigl( a_0 
+ a_1 Q^\dagger_1  
+ a_2 Q^\dagger_2 
+ a_3 Q^\dagger_1 Q^\dagger_2 \bigr) |0^{\text{zm}}\rangle}
{\sqrt{{|a_0|}^2+{|a_1|}^2 +{|a_2|}^2+{|a_3|}^2}}
\ ,
\label{ZeroModeState}
\end{align}
where the four $a_i$ are complex numbers, not all of them vanishing, we find 
\begin{align}
W^{(2,\overline2)}_{\text{zm}}
:= \langle\text{ZM}| 
\overline{\psi^{\text{zm}}} \, \psi^{\text{zm}} \overline{\psi^{\text{zm}}} \, \psi^{\text{zm}} 
|\text{ZM}\rangle
= \theta L^{-2}
\ , 
\label{eq:W2-zeromode}
\end{align}
where 
$\theta = \bigl({|a_0|}^2 +{|a_3|}^2 \bigr)
\bigl( {|a_0|}^2+{|a_1|}^2 +{|a_2|}^2+{|a_3|}^2 \bigr)^{-1}$. 
Note that $\theta \in [0,1]$, and when 
$|\text{ZM}\rangle = |0^{\text{zm}}\rangle$, we have $\theta=1$.

\subsubsection{Full field}

Consider now the full field~\eqref{eq:psi-osc+zm-split}, 
consisting of both the oscillator modes and the zero mode. 
We put the field in the state 
\begin{align}
|\widetilde{\text{ZM}}\rangle 
& := 
|\text{ZM}\rangle \otimes |0^{\text{osc}}\rangle
\notag
\\
& = 
\frac{\bigl( a_0 
+ a_1 Q^\dagger_1  
+ a_2 Q^\dagger_2 
+ a_3 Q^\dagger_1 Q^\dagger_2 \bigr)
\bigl( |0^{\text{zm}}\rangle \otimes |0^{\text{osc}}\rangle \bigr)}
{\sqrt{{|a_0|}^2+{|a_1|}^2 +{|a_2|}^2+{|a_3|}^2}}
\ ,
\label{eq:productstate}
\end{align}
We show in Appendix \ref{app:decoupling} that 
\begin{align}
W^{(2,\overline2)}_u(t,x; t', x') 
& := 
\langle \widetilde{\text{ZM}}| \, 
\overline{\psi}(t,x) \psi(t,x)
\overline{\psi}(t',x') \psi(t',x')
\, 
|\widetilde{\text{ZM}}\rangle 
\notag
\\
& = 
W^{(2,\overline2)}_{\text{osc}}(t,x; t', x') 
+ 
W^{(2,\overline2)}_{\text{mix}}(t,x; t', x') 
+ 
W^{(2,\overline2)}_{\text{zm}} 
\ , 
\label{eq:untwisted-totalW2}
\end{align}
where $W^{(2,\overline2)}_{\text{osc}}$ is given by~\eqref{eq:W2-osc-final}, 
$W^{(2,\overline2)}_{\text{zm}}$ is given by~\eqref{eq:W2-zeromode}, 
\begin{align}
W^{(2,\overline2)}_{\text{mix}} 
&= 
L^{-2}
\sum_{n=1}^{\infty}
\left( 
e^{-2 \pi i n (\Delta t+ \Delta x -i\epsilon)/L} 
+ e^{-2 \pi i n (\Delta t- \Delta x -i\epsilon)/L} 
\right) 
\notag
\\
&= L^{-2}
\left( 
\frac{1}{e^{2 \pi i(\Delta t+ \Delta x -i\epsilon)/L} -1}
+ \frac{1}{e^{2 \pi i(\Delta t- \Delta x -i\epsilon)/L} -1}
\right)
\ , 
\label{eq:W2-mix-final} 
\end{align}
and the subscript in $W^{(2,\overline2)}_u$ 
refers to the untwisted spin structure. 

The response of the detector is obtained from \eqref{RespFunct-def} 
with~\eqref{eq:untwisted-totalW2}. 
Note that the response again contains no infrared ambiguities. 
We may break the response as 
\begin{align}
\mathcal{F}_u(\Omega) 
= 
\mathcal{F}_u^{\text{osc}}(\Omega) 
+ 
\mathcal{F}_u^{\text{mix}}(\Omega)
+ 
\mathcal{F}_u^{\text{zm}}(\Omega)
\ , 
\label{eq:Funwisted-general}
\end{align}
where numerically efficient formulas for 
$\mathcal{F}_u^{\text{osc}}$ and $\mathcal{F}_u^{\text{mix}}$ are obtained from the sums in 
\eqref{eq:W2-osc-final} and~\eqref{eq:W2-mix-final}, 
\begin{subequations}
\label{eq:Funwisted-general-osc+mix}
\begin{align}
\mathcal{F}_u^{\text{osc}}(\Omega) 
&= 
\frac{2}{L^2}  
\sum_{n,m=1}^{\infty}
\int d\tau \, d \tau'  \, 
\chi(\tau)\chi(\tau')
\, e^{-i\Omega(\tau-\tau^\prime)}
\notag
\\[1ex]
&\hspace{3ex}
\times 
\exp \! \left(
-\frac{2 \pi i \left\{ (n+m)[t(\tau)-t(\tau^\prime) - i\epsilon]
+ (n-m)[(x(\tau)-x(\tau^\prime)] \right\}}{L}
\right) 
\ , 
\\
\mathcal{F}_u^{\text{mix}}(\Omega) 
&= 
\frac{2}{L^2}  
\sum_{n=1}^{\infty}
\int d\tau \, d \tau'  \, 
\chi(\tau)\chi(\tau')
\, e^{-i\Omega(\tau-\tau^\prime)}
\notag
\\[1ex]
&\hspace{6ex}
\times 
\exp \! \left(
-\frac{2 \pi i n[t(\tau)-t(\tau^\prime) - i\epsilon]}{L}
\right) 
\cos \! \left(
\frac{2 \pi n[x(\tau)-x(\tau^\prime)]}{L}
\right) 
\ , 
% \label{GeneralRFTS}
\end{align}
\end{subequations}
while 
\begin{align}
\label{ZMRF}
\mathcal{F}_u^{\text{zm}}(\Omega) 
&= 
\frac{\theta}{L^2} 
\int d\tau \, d \tau'  \, 
\chi(\tau)\chi(\tau')
\, e^{-i\Omega(\tau-\tau^\prime)}
%\notag
%\\
%& 
= 
\frac{\theta}{L^2} \, |\widehat{\chi}(\Omega)|^2
\ ,
\end{align}
where the hat denotes the Fourier transform, 
$\widehat\chi(\Omega) := \int_{-\infty}^{\infty} d\tau \, \chi(\tau) \, e^{-i\Omega\tau}$. 

\subsection{$L\to\infty$ limit\label{subsec:L-to-infty}}

In the limit $L\to\infty$, the final expressions 
in \eqref{General2-pointCFTS}, 
\eqref{eq:W2-osc-final}, 
\eqref{eq:W2-zeromode}
and 
\eqref{eq:W2-mix-final} 
show that 
both 
$W^{(2,\overline2)}_{t}(t,x; t', x')$ and $W^{(2,\overline2)}_u(t,x; t', x')$
approach the same limit, 
\begin{align}
W^{(2,\overline2)}_{t,u}(t,x; t', x')
\xrightarrow[L\to\infty]{} 
\frac{1}{2\pi^2 \bigl[(x-x')^2 - (t-t' - i\epsilon)^2 \bigr]}
\ , 
\label{eq:W2-Mink-limit}
\end{align}
which by \eqref{eq:CorrFunct-final-massless}
is equal to $W^{(2,\overline2)}(t,x; t', x')$ 
in the Minkowski vacuum in two-dimensional Minkowski spacetime. 
This is as expected: in the limit of large spatial circumference, 
the detector's response for either spin structure reduces to that 
in the Minkowski vacuum in Minkowski spacetime. 

%Note, by Theorem~\ref{thm:massless-fermion-to-scalar}, 
%that \eqref{eq:W2-Mink-limit}
%equals twice the Wightman function 
%of a massless scalar field in four-dimensional Minkowski spacetime. 

\section{Inertial and uniformly accelerated trajectories on the 
cylindrical spacetime\label{sec:sample-trajectories}}

In this section we consider inertial and uniformly 
accelerated detectors on the cylindrical spacetime of 
Section~\ref{sec:cylinder}.

\subsection{Inertial detector}

Consider a detector on the inertial worldline 
\begin{align}
t= \tau \cosh\beta\ , 
\hspace{4ex}
x = \tau\sinh\beta
\ ,
\label{eq:inertial-worldline}
\end{align}
where $\beta\in\BbbR$ is the rapidity with respect to the worldlines of constant~$x$.  
We take the switching function to be Gaussian, 
\begin{align}
\chi(\tau) 
= 
\frac{1}{\pi^{1/4}\sigma^{1/2}} \, e^{-\tau^2/(2\sigma^2)}
\ ,
\label{eq:gaussian-switching}
\end{align}
where the positive parameter $\sigma$ is the effective duration of the interaction. 
The normalisation is such that $\int_{-\infty}^{\infty}\chi^2(\tau) d\tau = 1$, and 
\begin{align}
\widehat{\chi}(\Omega) = 
\pi^{1/4}(2\sigma)^{1/2} \, e^{-\sigma^2\Omega^2/2} 
\ .
\end{align}

For the twisted field, \eqref{GeneralRFTS} gives 
\begin{align}
\mathcal{F}_t(\Omega) 
= 
\frac{4\pi^{1/2}\sigma}{L^2}
\sum_{n,m=0}^{\infty}
\exp \! \left[-\sigma^2 
\left( \Omega + \frac{2\pi\bigl((n+\tfrac12) e^{\beta} + (m+\tfrac12)e^{-\beta}\bigr)}{L} \right)^{\!2} \, 
\right]
\ . 
\label{eq:RFTwisted}
\end{align}
For the untwisted field, 
\eqref{eq:Funwisted-general}, 
\eqref{eq:Funwisted-general-osc+mix} and \eqref{ZMRF} give 
\begin{align}
\mathcal{F}_u(\Omega) 
& = 
\frac{4\pi^{1/2}\sigma}{L^2}
\sum_{n,m=1}^{\infty}
\exp \! \left[-\sigma^2 
\left( \Omega + \frac{2\pi (n e^{\beta} + me^{-\beta})}{L} \right)^{\!2} \, 
\right]
\notag
\\ 
& \hspace{3ex}
+ 
\frac{2\pi^{1/2}\sigma}{L^2}
\sum_{n=1}^{\infty}
\left\{
\exp \! \left[-\sigma^2 
\left( \Omega + \frac{2\pi n e^{\beta}}{L} \right)^{\!2} \, 
\right] 
+ \exp \! \left[-\sigma^2 
\left( \Omega + \frac{2\pi n e^{-\beta}}{L} \right)^{\!2} \, 
\right]
% ^{\vphantom{X}}
\right\}
\notag
\\ 
& \hspace{3ex}
+ \frac{2\pi^{1/2}\sigma\theta}{L^2} \, e^{-\sigma^2\Omega^2} 
\ . 
\label{eq:RFUntwisted}
\end{align}

Three comments are in order. 

First, consider the limit of long detection, $\sigma\to\infty$, 
with the other parameters fixed. In this limit, 
$\mathcal{F}_t$ and $\mathcal{F}_u$ each reduce to a series of delta-peaks, 
\begin{subequations}
\begin{align}
\mathcal{F}_t(\Omega) 
&= 
\frac{4\pi}{L^2}\sum_{n,m=0}^{\infty}
\delta \! 
\left( \Omega + \frac{2\pi\bigl((n+\tfrac12) e^{\beta} + (m+\tfrac12)e^{-\beta}\bigr)}{L} \right)
\ , 
\label{eq:erl20}
\\
\mathcal{F}_u(\Omega) 
&=  
\frac{4\pi}{L^2}\sum_{n,m=1}^{\infty}
\delta \! 
\left( \Omega + \frac{2\pi (n e^{\beta} + me^{-\beta})}{L} \right)
\notag
\\
& \hspace{3ex}
+ 
\frac{2\pi}{L^2}\sum_{n=1}^{\infty}
\left[
\delta \! 
\left( \Omega + \frac{2\pi n e^{\beta}}{L} \right)
+ 
\delta \! 
\left( \Omega + \frac{2\pi n e^{-\beta}}{L} \right)
% ^{\vphantom{x^x}}
\right]
\notag
\\
& \hspace{3ex}
+ 
\frac{2\pi \theta}{L^2} 
\, \delta ( \Omega ) 
\ ,
\label{eq:erl16}
\end{align}
\end{subequations}
where $\delta$ is Dirac's delta function. 
The Doppler shift factors $e^{\pm\beta}$  
show that the peaks in $\mathcal{F}_t$
correspond to the creation of a pair of field excitations, 
one left-moving and the other right-moving. 
The peaks in $\mathcal{F}_u$ are similar but also contain the 
special cases where one or both of the field excitations are in the zero mode, 
with vanishing energy. 
That the excitations occur in pairs is a consequence 
of the quadratic interaction Hamiltonian 
$H_{\text{int}}$~\eqref{InterHamilt}.  
By contrast, the peaks for a 
detector coupled linearly to a scalar field 
\cite{Martin-Martinez:2014qda} correspond to emission 
of just single field quanta. 

Second, consider the ultrarelativistic velocity limit, $|\beta|\to \infty$, 
with the other parameters fixed. 
$\mathcal{F}_t$~vanishes in this limit, 
exponentially in~$e^{|\beta|}$: 
the physical reason is that the detector would need to excite 
field quanta in pairs and one member of each pair is necessarily 
highly blueshifted in the detector's local rest frame. 
For $\mathcal{F}_u$, however, 
one of the single sums in \eqref{eq:RFUntwisted} does not vanish in this limit, 
and estimating the sum by an integral gives 
\begin{align}
\mathcal{F}_u(\Omega) = \frac{e^{|\beta|}}{2L} 
\bigl[ \erfc(\sigma\Omega) + o(1) \bigr]
\ \ \ \text{as $|\beta| \to \infty$}, 
\end{align}
where $\erfc$ is the error complement function~\cite{dlmf}. 
The physical interpretation is that at ultrarelativistic velocities 
the detector has an exponentially large probability to generate field excitation pairs 
in which one excitation is highly redshifted with respect 
to the detector's local rest frame and the other excitation is a zero mode. 
This phenomenon has no counterpart for a 
detector coupled linearly to a scalar field~\cite{Martin-Martinez:2014qda}. 

\begin{figure}[p]
\centering
\subfigure[$L\mathcal{F}_t$ for $\beta=0$]{%
\includegraphics[width=0.48\textwidth]{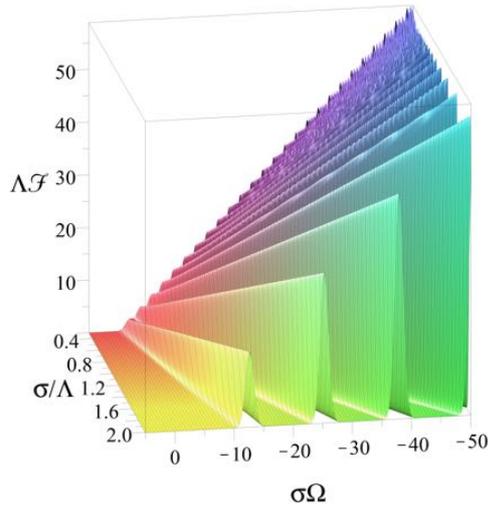} % low res
\label{TSRFplotbeta0}}
\subfigure[$L\mathcal{F}_u$ for $\beta=0$]{%
\includegraphics[width=0.48\textwidth]{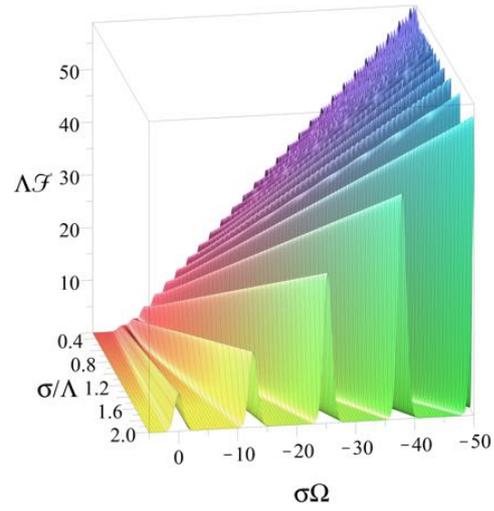} % low res
\label{USRFplotbeta0}}
\subfigure[$L\mathcal{F}_t$ for $\beta=1$]{%
\includegraphics[width=0.48\textwidth]{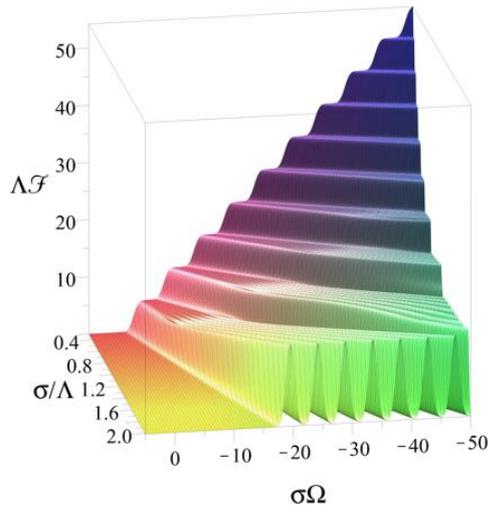} % low res
\label{TSRFplotbeta1}}
\subfigure[$L\mathcal{F}_u$ for $\beta=1$]{%
\includegraphics[width=0.48\textwidth]{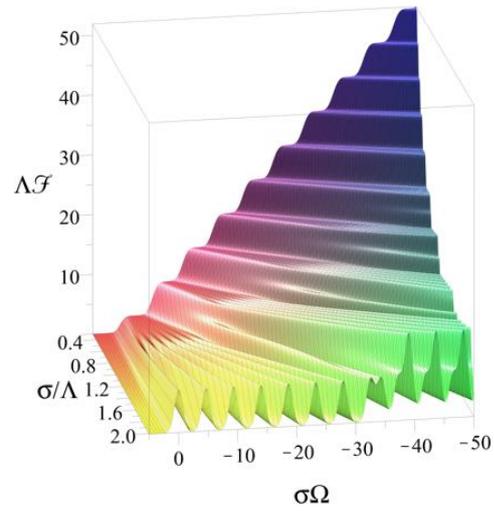} % low res
\label{USRFplotbeta1}}
\caption{Perspective plots of $L\mathcal{F}_t$ and $L\mathcal{F}_u$ 
for the inertial detector in terms of the dimensionless variables 
$\sigma\Omega$ and $\sigma/L$, 
for $\beta=0$ and $\beta=1$. In $L\mathcal{F}_u$ we have set $\theta=1$. 
$\Lambda$~in the axis labels stands for~$L$.}
\label{TWUN1} 
\end{figure}

\begin{figure}[p]
\centering
\subfigure[$L\mathcal{F}_t$ for $\beta=0.5$]{%
\includegraphics[width=0.48\textwidth]{low_L_F_twisted_beta_0p5.pdf} % low res
\label{TSRFplotbeta0.5}}
\subfigure[$L\mathcal{F}_u$ for $\beta=0.5$]{%
\includegraphics[width=0.48\textwidth]{low_L_F_untwisted_beta_0p5.pdf} % low res
\label{USRFplotbeta0.5}}
\\[5ex]
\subfigure[Cross-section of \subref{TSRFplotbeta0.5} at $\sigma/L =1$.]{%
\includegraphics[width=0.45\textwidth]{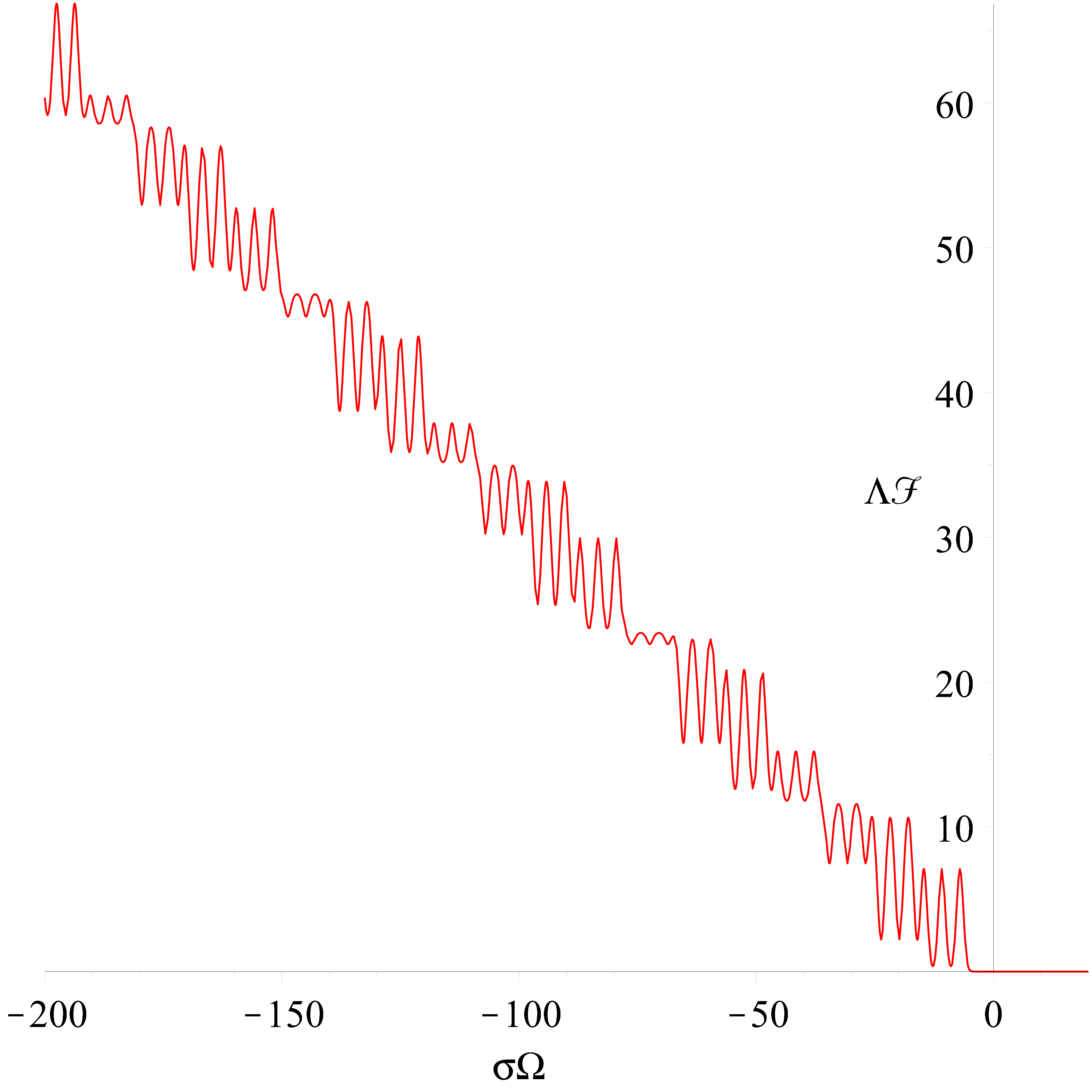}
\label{TSRF2Dplotbeta0.5}}
\quad 
\subfigure[Cross-section of \subref{USRFplotbeta0.5} at $\sigma/L =1$.]{%
\includegraphics[width=0.45\textwidth]{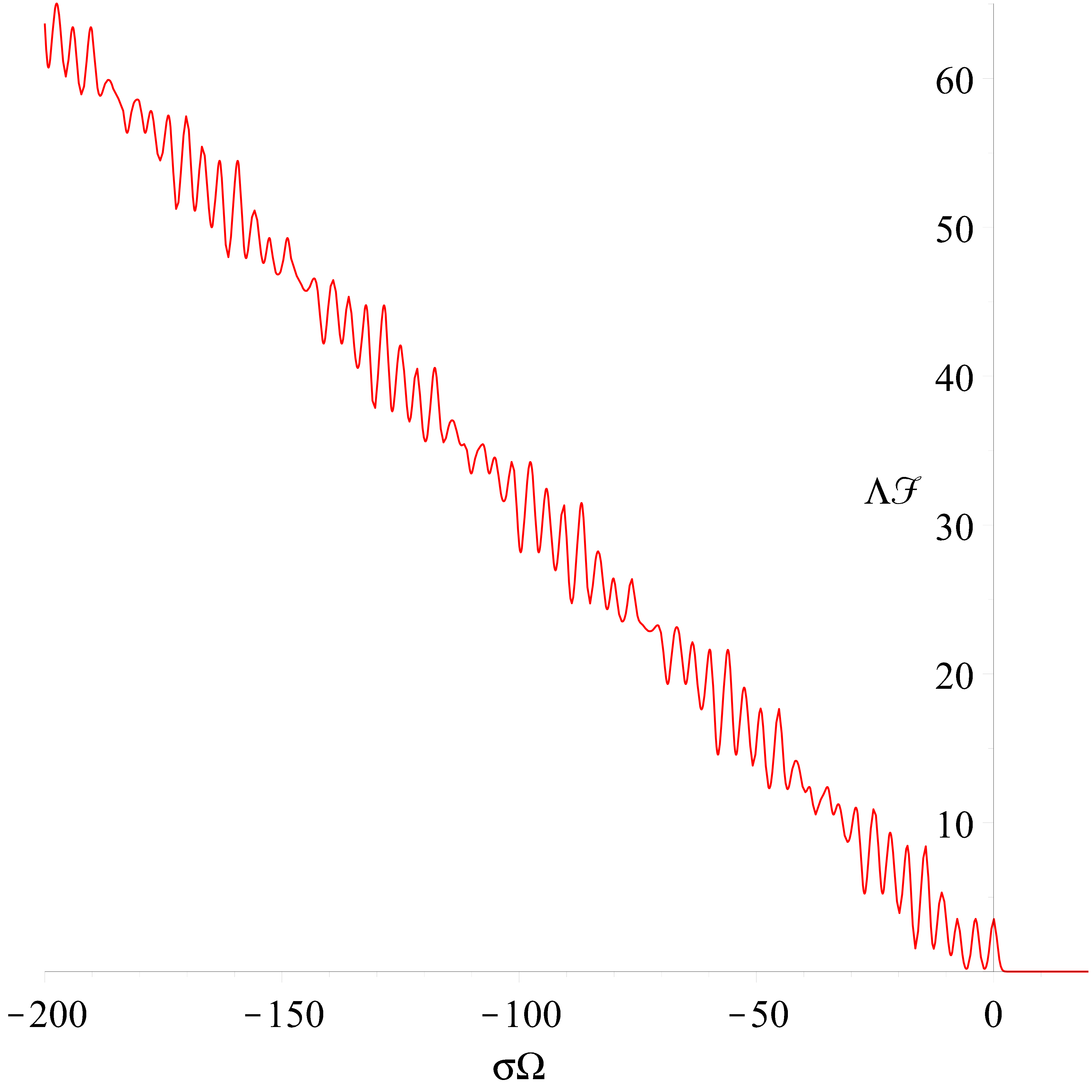}
\label{USRF2Dplotbeta0.5}}
\caption{As in Figure \ref{TWUN1} but with $\beta=0.5$, 
showing also cross-sections at $\sigma/L =1$ which reveal finer detail.} 
\label{TWUN2}
\end{figure} 

\begin{figure}[p]
\centering
\subfigure[$a^{-1}\mathcal{F}_t$]{%
\includegraphics[width=0.48\textwidth]{low_F_twisted3D.pdf} % low res
\label{USRF_acc1}}
\subfigure[$a^{-1}\mathcal{F}_u$]{%
\includegraphics[width=0.48\textwidth]{low_F_untwisted3D.pdf} % low res
\label{TSRF_acc1}}
\\[5ex]
\subfigure[Cross-section of \subref{USRF_acc1} at $aL=1$.]{%
\includegraphics[width=0.45\textwidth]{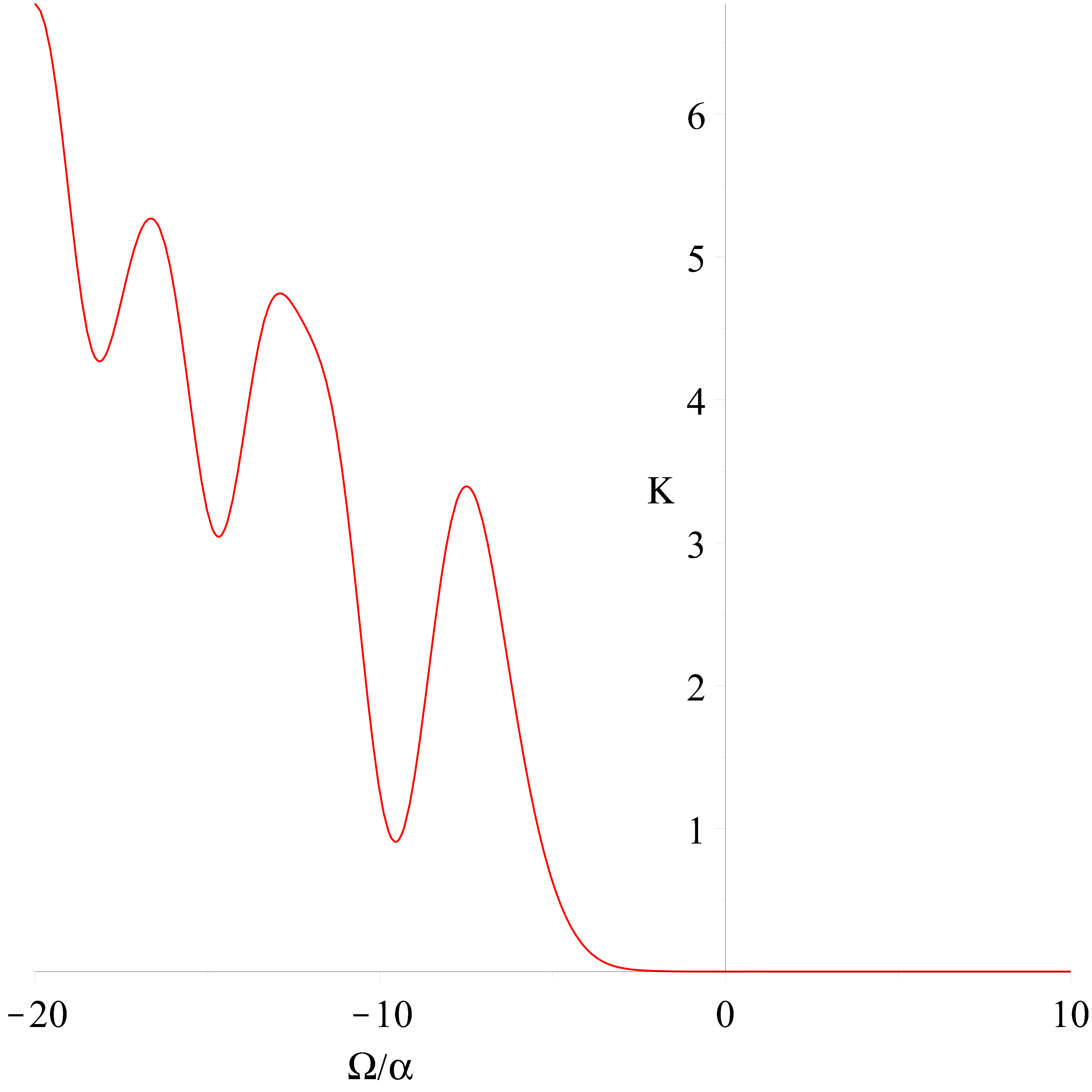}
\label{USRF_acc2}}
\quad
\subfigure[Cross-section of \subref{TSRF_acc1} at $aL=1$.]{%
\includegraphics[width=0.45\textwidth]{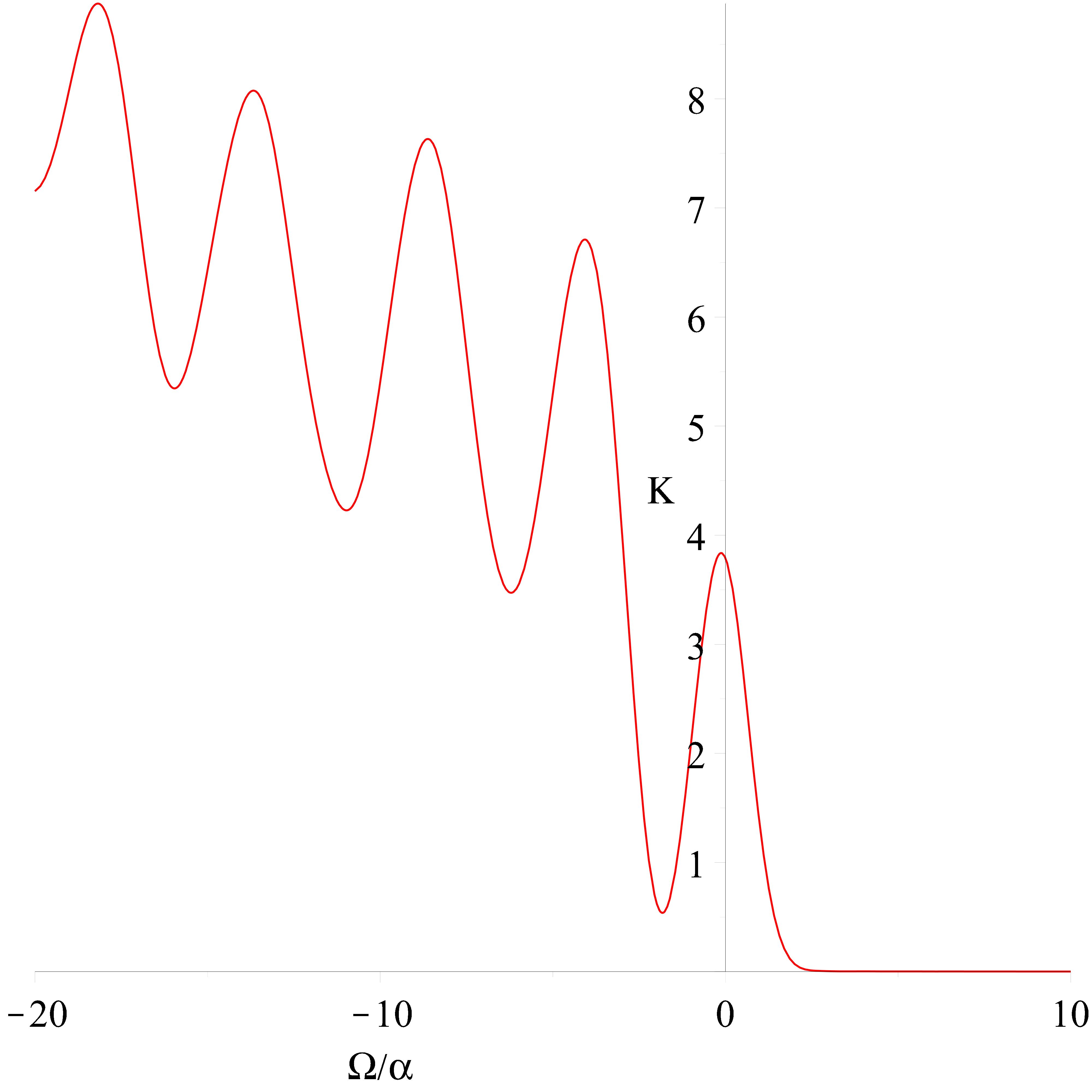}
\label{TSRF_acc2}}
\caption{Perspective plots of 
$a^{-1}\mathcal{F}_t$ and $a^{-1}\mathcal{F}_u$ 
for the uniformly accelerated detector 
in terms of the dimensionless variables 
$aL$ and $\Omega/a$, for $\tau_0=0$, 
and cross-sections at $aL=1$. 
In $a^{-1}\mathcal{F}_u$ we have set $\theta=1$. 
$K$, $\Lambda$ and $\alpha$ in the axis labels stand respectively for 
$a^{-1}\mathcal{F}$, $L$ and~$a$.}
\label{TWUN3}
\end{figure}

Third, consider the large circumference limit, 
$L\to\infty$, with the other parameters fixed. 
As noted in subsection~\ref{subsec:L-to-infty}, 
in this limit both $\mathcal{F}_t$ and $\mathcal{F}_u$
approach the response of an inertial detector 
in Minkowski vacuum in $(1+1)$-dimensional 
Minkowski spacetime, evaluated in 
Appendix~\ref{app:stationary1+1}, with the result
\begin{align}
\mathcal{F}_{t,u}(\Omega) \xrightarrow[L\to\infty]{} 
%\frac{4\pi^{1/2}}{\sigma}\int_{0}^{\infty} dv \, v  
%\exp\bigl[-(2\pi v+\sigma\Omega)^2 \bigr] 
%\notag 
%\\[1ex]
%&=
\frac{1}{2\pi\sigma}\left(\frac{e^{-\sigma^2\Omega^2}}{\pi^{1/2}} -\sigma\Omega \, 
\erfc(\sigma\Omega) \right)
\ . 
\label{eq:erl4}
\end{align}
In the limit $\sigma\to\infty$, 
\eqref{eq:erl4} reduces to 
\begin{align}
\mathcal{F}(\Omega) = -\frac{\Omega \Theta(-\Omega)}{\pi}
\ , 
\label{eq:erl5}
\end{align}
where $\Theta$ is the Heaviside function. 
Formula \eqref{eq:erl5} equals twice the response of
an inertial Unruh-DeWitt detector coupled linearly to a scalar field 
in four-dimensional Minkowski space 
in the long interaction limit~\cite{Birrell:1982ix}, 
as must be the case by Theorem~\ref{thm:massless-fermion-to-scalar}. 

Plots of $L\mathcal{F}_t$ and $L\mathcal{F}_u$ 
as a function of the dimensionless variables $\sigma\Omega$ 
and $\sigma/L$ are shown in Figures
\ref{TWUN1} and~\ref{TWUN2}.

\subsection{Uniformly accelerated detector} 

Consider a detector on the uniformly accelerated worldline 
\begin{align}
t = a^{-1}\sinh(a\tau)
\ , 
\hspace{4ex}
x = a^{-1}\cosh(a\tau)
\ ,
\label{eq:rindler-trajectory}
\end{align}
where the positive parameter $a$ is the proper acceleration. 
As this trajectory is not stationary on the cylinder, 
we now consider the Gaussian switching function 
\begin{align}
\chi_{\tau_0}(\tau) = \frac{1}{\pi^{1/4}\sigma^{1/2}} \, 
e^{-(\tau-\tau_0)^2/(2\sigma^2)}
\ ,
\label{eq:chi-gauss-taunought}
\end{align}
where 
% $\sigma > 0$ and 
the new real-valued parameter 
$\tau_{0}$ specifies the moment about which 
$\chi_{\tau_0}$ is peaked.

For the twisted field, \eqref{GeneralRFTS} gives 
\begin{align}
\mathcal{F}_t(\Omega) 
= \frac{2}{\pi^{1/2}{(aL)}^2\sigma}
\sum_{n,m=0}^{\infty}
\left|I^t_{nm}\right|^2
\ ,
\label{eq:AcceleratedRF1}
\end{align}
where
\begin{align}
I^t_{nm} = \int_{0}^{\infty}\frac{dy}{y}  
\exp \! \left[ -\frac{{(\ln y - a\tau_0)}^2}{2\sigma^2a^2}
- \frac{i\Omega \ln y}{a} - \frac{2\pi i}{aL}
\bigl[(n+\tfrac12)y-(m+\tfrac12) y^{-1} \bigr] \right]
\ . 
\label{eq:It-def}
\end{align}
For the untwisted field, 
\eqref{eq:Funwisted-general}, 
\eqref{eq:Funwisted-general-osc+mix} and \eqref{ZMRF} give 
\begin{align}
\mathcal{F}_u(\Omega) 
& = \frac{2}{\pi^{1/2}{(aL)}^2\sigma}
\sum_{n,m=1}^{\infty}
\left|I^u_{nm}\right|^2
\notag
\\ 
& \hspace{3ex}
+ \frac{1}{\pi^{1/2}{(aL)}^2\sigma}
\sum_{n=1}^{\infty}
\left( 
\left|J^+_{n}\right|^2 + \left|J^-_{n}\right|^2
\right) 
\notag
\\ 
& \hspace{3ex}
+ \frac{2\pi^{1/2}\sigma\theta}{L^2} \, e^{-\sigma^2\Omega^2} 
\ ,
\label{eq:AcceleratedRF2}
\end{align}
where
\begin{subequations}
\begin{align}
I^u_{nm} 
&= 
\int_{0}^{\infty}\frac{dy}{y}  
\exp \! \left[ -\frac{{(\ln y - a\tau_0)}^2}{2\sigma^2a^2}
 - \frac{i\Omega \ln y}{a} - \frac{2\pi i}{aL}
\bigl(ny-my^{-1}\bigr) \right]
\ , 
\\
J^+_{n} 
&= 
\int_{0}^{\infty}\frac{dy}{y}  
\exp \! \left[ -\frac{{(\ln y - a\tau_0)}^2}{2\sigma^2a^2}
 - \frac{i\Omega \ln y}{a} - \frac{2\pi n i}{aL} \, y \right]
\ , 
\\
J^-_{n} 
&= 
\int_{0}^{\infty}\frac{dy}{y}  
\exp \! \left[ -\frac{{(\ln y - a\tau_0)}^2}{2\sigma^2a^2}
 - \frac{i\Omega \ln y}{a} + \frac{2\pi n i}{aL} \, y^{-1} \right]
\ . 
\end{align}
\end{subequations} 

As the detector's worldline is not stationary, 
analytic investigation of $\mathcal{F}_t$ and $\mathcal{F}_u$ 
in the limit of large $\sigma$ and in the limit of large $\tau_0$ is not straightforward. 
In the limit of large circumference, however, we recall from 
subsection \ref{subsec:L-to-infty} 
that both $\mathcal{F}_t$ and $\mathcal{F}_u$
approach the response in Minkowski vacuum, evaluated in 
Appendix~\ref{app:stationary1+1}, with the result 
\begin{align}
\mathcal{F}_{t,u}(\Omega)&\xrightarrow[L\to\infty]{} 
\frac{a e^{-\pi\Omega/a}}{4\pi^2}
\int_{-\infty}^\infty
\frac{dr}{\cosh^{2\!} r}
\, 
\exp \! \left( -\frac{{(r - i\pi/2)}^2}{\sigma^2a^2}
-\frac{2i\Omega r}{a} \right)
\ . 
\label{eq:erl11}
\end{align}
In the limit $\sigma\to\infty$, \eqref{eq:erl11} 
reduces by formula 3.982.1 in \cite{grad-ryzh} to 
the Planckian distribution in 
the Unruh temperature $a/(2\pi)$, 
\begin{align}
\mathcal{F}(\Omega) =
\frac{\Omega}{\pi \bigl( e^{2\pi \Omega/a}-1 \bigr)}
\ . 
\label{eq:erl5planck}
\end{align}
Formula \eqref{eq:erl5planck} equals twice the response of
a uniformly accelerated Unruh-DeWitt detector coupled linearly to a scalar field 
in four-dimensional Minkowski space 
in the long interaction limit~\cite{Birrell:1982ix}, 
as must be the case by Theorem~\ref{thm:massless-fermion-to-scalar}. 

Plots of $a^{-1}\mathcal{F}_t(\Omega)$ and $a^{-1}\mathcal{F}_u(\Omega)$ as a function of 
the dimensionless variables $aL$ and $\Omega/a$ are shown in Figure~\ref{TWUN3}.

% \newpage 

\section{Conclusions\label{sec:conclusions}}

We have analysed the response 
of a spatially pointlike Unruh-DeWitt detector 
coupled linearly to the scalar density of a massless Dirac field 
in Minkowki spacetimes in dimension $d\ge2$ 
and on the $(1+1)$-dimensional flat static cylinder, 
allowing the detector's motion to remain arbitrary and 
allowing the detector to be switched on and off in an arbitrary smooth way. 
Working within first-order perturbation theory, 
we regularised the interaction by dropping 
an additive term that is technically ill-defined but formally proportional to the 
field's mass~\cite{Langlois:2005nf,Langlois:2005if}. 

In $d$-dimensional Minkowski, with the field in its Fock vacuum, 
we found that the response is identical to that of a detector 
coupled linearly to a massless scalar field in $2d$ spacetime dimensions. 
For a uniformly linearly accelerated detector, 
this implies that the long time limit of the response exhibits
the Unruh effect with a Planckian frequency dependence factor, for all~$d$. 
While the Rindler power spectrum of the Dirac 
field is known to have a Planckian factor for odd $d$ but a 
Fermi-Dirac factor for even~$d$~\cite{Takagi:1986kn}, 
the detector's response is Planckian for all $d$ because the 
response is not proportional to the Rindler power spectrum 
but to the convolution of the Rindler power spectrum with itself. 

In the special case of two-dimensional Minkowski, we saw that 
the detector's response has no infrared ambiguity. 
In this respect our detector differs from the 
detector coupled linearly to a massless scalar field, 
where in two dimensions the response is ambiguous due to the 
infrared ambiguity of the Wightman function~\cite{Decanini:2005eg}. 

On the $(1+1)$-dimensional flat static cylinder, 
we found that the response distinguishes the Fock vacua of the field's oscillator modes for 
periodic and antiperiodic spin structures, 
and the zero mode that occurs for the periodic spin structure contributes to the response 
in a way that depends on zero mode's initial state. 
We also provided a selection of analytic and numerical 
results for inertial and uniformly accelerated
trajectories on the cylinder, recovering the $d=2$ Minkowski 
results in the limit of large circumference. 

While we have focused the present paper on static flat spacetimes and 
to quantum states that are invariant under translations in the Killing time, 
there would be scope for examining the detector coupled to the 
Dirac field in more general spacetimes and for more general quantum states, 
including collapsing star spacetimes \cite{hawking}
and their flat ``moving mirror'' counterparts~\cite{Birrell:1982ix,Juarez-Aubry:2014jba}, 
or spatially homogeneous cosmologies, where Dirac's equation can be solved 
by separation of variables~\cite{Duncan:1977fc}. For example, 
if a cosmological spacetime has a de~Sitter era, exactly or approximately, 
how does the detector register the associated 
Gibbons-Hawking temperature~\cite{gibb-haw:dS}? 
We leave these questions subject to future work.

\section*{Acknowledgments}

We thank Benito Ju\'arez-Aubry and Eduardo Mart\'in-Mart\'inez for helpful discussions. 
JL is supported in part by STFC (Theory Consolidated Grant ST/J000388/1). 
VT is supported in part by U.S. Department of Education under U.S. Federal Student Aid.

\appendix

\section{Gamma matrices and basis spinors\label{AppenGamma}}

In this appendix we record relevant properties of the gamma-matrices 
and the massive basis spinors in spacetime dimension $d\ge2$. 
More detail can be found in~\cite{parker-toms-book}. 

The gamma-matrices $\gamma^\mu$, $\mu = 0,1,\ldots,d-1$, 
%are defined in terms of 
%irreducible representations 
%of the Clifford algebra Cliff$(1,d-1)$. 
are $N_d\times N_d$ matrices with 
\begin{align}
\label{DiraSpN1}
N_d =
\begin{cases}
2^{d/2} & \text{for $d$ even}, 
\\
2^{(d-1)/2} & \text{for $d$ odd}, 
\end{cases}
\end{align}
satisfying 
\begin{align}
\label{ref:AntiComm}
\lbrace\gamma^\mu, \gamma^\nu\rbrace = 2\eta^{\mu\nu}
\ , 
\end{align}
where on the right hand side we have suppressed the identity matrix 
$I_{N_d\times N_d}$. $\gamma^0$ is Hermitian, 
$\gamma^1, \ldots , \gamma^{d-1}$ are anti-Hermitian, 
$\Tr(\gamma^\mu)=0$, and $\Tr(\gamma^\mu \gamma^\nu)=N_d \eta^{\mu\nu}$. 

%We employ the Weyl representation, in which 
%\begin{align}
%\label{Gamma0_N}
%\gamma^0 = 
%\begin{pmatrix}
%0 & I_{N_d/2\times N_d/2}\\
%I_{N_d/2 \times N_d/2} &  0
%\end{pmatrix}
%\end{align}
%and $\Tr(\gamma^\mu)=0$. Note that $\Tr(\gamma^\mu \gamma^\nu)=N_d \eta^{\mu\nu}$. 
%For $d=2$, we have 
%$\gamma^0 = \sigma_1 = \left(\begin{smallmatrix
%$\gamma^1 = i\sigma_2 = \left(\begin{smallmatrix}0&1\\-1&0\end{smallmatrix}\right)$, 
%where $\sigma_1$ and $\sigma_2$ are two of the Pauli sigma matrices. 

Let $u^{(s)}(\textbf{0})$ and $v^{(s)}(\textbf{0})$ be 
eigenspinors of $\gamma^0$ such that 
\begin{subequations}
\begin{align}
\gamma^0 u^{(s)} (\textbf{0}) &= u^{(s)} (\textbf{0}) 
\ , 
\\
\gamma^0 v^{(s)} (\textbf{0}) &= - v^{(s)} (\textbf{0}) 
\ , 
\end{align}
\end{subequations}
with the orthonormality conditions 
\begin{subequations}
\begin{align}
u^{(s)\dagger}(\textbf{0})u^{(s^\prime)}(\textbf{0})  
& = v^{(s)\dagger}(\textbf{0})v^{(s^\prime)}(\textbf{0}) 
= 
2 m\delta^{ss^\prime}
\ , 
\\
u^{(s)\dagger}(\textbf{0})v^{(s^\prime)}(\textbf{0})
& = 0
\ , 
\end{align}
\end{subequations}
where the helicity index 
$s$ takes the values $s = 1,\dots,N_d/2$. 
The spinors $u^{(s)}(\textbf{k})$ and $v^{(s)}(\textbf{k})$ are defined by 
\begin{subequations}
\begin{align}
u^{(s)}(\textbf{k})
& =
\frac{\gamma^\mu k_\mu + m}{\sqrt{2 m (k^0 + m)}} 
\, 
u^{(s)}(\textbf{0})
\ , 
\\ 
v^{(s)}(\textbf{k})
& =
\frac{-\gamma^\mu k_\mu + m}{\sqrt{2 m (k^0 + m)}} 
\, 
v^{(s)}(\textbf{0})
\ , 
\end{align}
\end{subequations}
where
$k^0 = 
% \omega_{\textbf{k}} = 
\bigl(\textbf{k}^2+m^2\bigr)^{1/2}$, 
and they satisfy 
\begin{subequations}
\begin{align}
(\gamma^\mu k_\mu - m) u^{(s)}(\textbf{k}) =0
\ , 
\\
(\gamma^\mu k_\mu + m) v^{(s)}(\textbf{k}) =0
\ . 
\end{align}
\end{subequations}
The orthonormality conditions are 
\begin{subequations}
\begin{align}
u^{(s)\dagger}(\textbf{k})u^{(s^\prime)}(\textbf{k})  
& = v^{(s)\dagger}(\textbf{k})v^{(s^\prime)}(\textbf{k}) 
= 
2 k^0\delta^{ss^\prime}
\ , 
\\ 
\overline{u}^{(s)}(\textbf{k})u^{(s^\prime)}(\textbf{k})
& = - \overline{v}^{(s)}(\textbf{k})v^{(s^\prime)}(\textbf{k})
= 
2 m \delta^{ss^\prime}
\ , 
\\
\overline{u}^{(s)}(\textbf{k})v^{(s^\prime)}(\textbf{k})
& = 0
\ , 
\end{align}
\end{subequations}
and the completeness identities are 
\begin{subequations}
\begin{align}
\sum_s u_a^{(s)}(\textbf{k}) \overline{u}_b^{(s)}(\textbf{k})
=(\gamma^\mu k_\mu + m)_{ab}
\ , 
\\ 
\sum_s v_a^{(s)}(\textbf{k}) \overline{v}_b^{(s)}(\textbf{k})
=(\gamma^\mu k_\mu - m)_{ab}
\ . 
\end{align}
\end{subequations}

\section{$W^{(2,\overline2)}$
in Minkowski vacuum\label{AppenB}}

In this appendix we write out the correlation function $W^{(2,\overline2)}(x,y)$ 
\eqref{eq:W2-gen-def}
in Minkowski spacetime in the Minkowski vacuum $| 0 \rangle$ 
in terms of the Wightman functions $S^\pm(x,y)$~\eqref{eq:bothWigFunct}. 
We treat the singular expression $S^-(x,x)$ here as a formal algebraic symbol 
but will address its interpretation in the main text. 

Setting $|\Psi_{0}\rangle = | 0 \rangle$, 
\eqref{eq:W2-gen-def} gives 
\begin{align}
\label{AppCorrFunct}
W^{(2,\overline2)} (x, y) 
& = 
\langle 0 
|\overline{\psi}(x) \psi(x)\overline{\psi}(y)\psi(y)
| 0 \rangle
\notag 
\\
& = 
\langle 0 
|\overline{\psi}_a(x) \psi_a(x)\overline{\psi}_b(y)\psi_b(y)
| 0 \rangle
\ , 
\end{align}
where each repeated spinor index is summed over. 

We use the decomposition 
\begin{align}
\overline{\psi}(x) \psi(x) 
& = 
N\bigl[\overline{\psi}(x) \psi(x)\bigr] 
+ \bigl\lbrace \overline{\psi}^+_a (x),\psi^-_a(x)\bigr\rbrace
\notag 
\\
& = 
N\bigl[\overline{\psi}(x) \psi(x)\bigr] 
- \Tr \bigl[ S^-(x,x) \bigr] 
\ , 
\label{eq:psibarpsi-decomp-normal}
\end{align}
where $N$ stands for the Wick normal product of a fermionic field, 
\begin{align}
N\bigl[\overline{\psi}(x) \psi(x)\bigr] 
:= 
\overline{\psi}^+_a(x) \psi^+_a(x)
+ \overline{\psi}^-_a(x) \psi^-_a(x)
+ \overline{\psi}^-_a(x) \psi^+_a(x) 
- \psi^-_a(x)\overline{\psi}^+_a(x)
\ , 
\label{eq:N-wicknormal}
\end{align}
and the last step in \eqref{eq:psibarpsi-decomp-normal} uses~\eqref{eq:bothWigFunct}. 
From \eqref{eq:psi-andbar-pm} we have 
\begin{subequations}
\label{eq:psis-annih-vac}
\begin{align}
\psi^+ | 0 \rangle 
&= 0 =  
\langle 0 |\psi^-
\ ,
\\
\overline{\psi}^+ | 0 \rangle 
&= 0 =  
\langle 0 | \overline{\psi}^-
\ , 
\end{align}
\end{subequations}
which shows that $\langle 0 | N\bigl[\overline{\psi}(x) \psi(x)\bigr] | 0 \rangle =0$. 
As $\Tr \bigl[ S^-(x,x) \bigr]$ is proportional to the identity operator in the Fock space, 
we hence have 
\begin{align}
W^{(2,\overline2)} (x, y) 
= \langle 0 | 
N\bigl[\overline{\psi}(x) \psi(x)\bigr]
N\bigl[\overline{\psi}(y) \psi(y)\bigr] 
| 0 \rangle
+ \Tr \bigl[ S^-(x,x) \bigr] 
\Tr \bigl[ S^-(y,y) \bigr] 
\ .
\label{eq:w2-decom-interm1}
\end{align}

For the first term in \eqref{eq:w2-decom-interm1}, we obtain 
\begin{align}
\langle 0 | 
N\bigl[\overline{\psi}(x) \psi(x)\bigr] 
N\bigl[\overline{\psi}(y) \psi(y)\bigr]
| 0 \rangle 
& =
\langle 0 | 
\overline{\psi}^+_a(x) \psi^+_a(x) 
\overline{\psi}^-_b(y) \psi^-_b(y) 
| 0 \rangle
\notag
\\
& =
\langle 0 | 
\overline{\psi}^+_a(x) 
\psi^-_b(y) 
\psi^+_a(x) 
\overline{\psi}^-_b(y) 
| 0 \rangle
\notag
\\
& =
\langle 0 | 
\bigl\lbrace \overline{\psi}^+_a(x) , \psi^-_b(y) \bigr\rbrace 
\bigl\lbrace \psi^+_a(x) , \overline{\psi}^-_b(y) \bigr\rbrace 
| 0 \rangle
\notag
\\
& = 
S^-_{ba}(y,x)
S^+_{ab}(x,y)
\notag
\\
& = 
\Tr \bigl[ S^+(x,y) S^-(y,x) \bigr]
\ , 
\end{align}
first using \eqref{eq:psis-annih-vac}, 
then 
anticommuting $\psi^-_b(y)$ past 
$\overline{\psi}^-_b(y)$ and 
$\psi^+_a(x)$, 
then using again \eqref{eq:psis-annih-vac} to 
replace 
$\overline{\psi}^+_a(x) \psi^-_b(y)$ 
and  
$\psi^+_a(x) \overline{\psi}^-_b(y)$ 
by anticommutators, and finally using the definition \eqref{eq:bothWigFunct}
of the Wightman functions~$S^\pm$ and cyclicity of the trace. 

Collecting, we have 
\begin{align}
W^{(2,\overline2)} (x, y) 
& = 
\Tr \bigl[ S^+(x,y) S^-(y,x) \bigr]
+ \Tr \bigl[ S^-(x,x) \bigr] 
\Tr \bigl[ S^-(y,y) \bigr] 
%\notag
%\\
%& = 
%\Tr \bigl[ S^+(x,y) S^-(y,x) \bigr]
%+ \Tr \bigl[ S^-(0,0) \bigr] 
%\Tr \bigl[ S^-(0,0) \bigr] 
\ , 
\label{AppCorrFunct-final}
\end{align}
from which \eqref{eq:CorrFunct-final-fromapp} follows 
using the translational invariance of $S^-(x,y)$.

\section{$W^{(2,\overline2)}$ of the untwisted massless Dirac field\label{app:decoupling}}

In this appendix we justify formula \eqref{eq:untwisted-totalW2}
for $W^{(2,\overline2)}$ of the untwisted massless 
Dirac field on the $(1+1)$-dimensional cylinder. 

Starting with \eqref{eq:untwisted-totalW2}, 
inserting the split \eqref{eq:psi-osc+zm-split}
and noting that terms with 
an unequal number of $\overline{\psi^{\text{osc}}}$s and $\psi^{\text{osc}}$s vanish, we obtain 
\begin{align}
W^{(2,\overline2)} (t,x; t', x') 
& = 
\langle \widetilde{\text{ZM}}| \, 
\overline{\psi}_a(t,x) \psi_a(t,x)
\overline{\psi}_b(t',x') \psi_b(t',x')
\, |\widetilde{\text{ZM}}\rangle 
\notag
\\
&= \Xi_1 + \Xi_2 + \Xi_3 + \Xi_4 + \Xi_5 + \Xi_6
\ , 
\end{align}
where 
\begin{subequations}
\begin{align}
\Xi_1 
& = 
\langle \widetilde{\text{ZM}}| \, 
\overline{\psi_a^{\text{osc}}}(t,x) \psi^{\text{osc}}_a(t,x)
\overline{\psi_b^{\text{osc}}}(t',x') \psi^{\text{osc}}_b(t',x')
\, |\widetilde{\text{ZM}}\rangle 
\ , 
\\
\Xi_2
& = 
\langle \widetilde{\text{ZM}}| \, 
\overline{\psi_a^{\text{osc}}}(t,x) \psi^{\text{osc}}_a(t,x)
\overline{\psi_b^{\text{zm}}} \psi^{\text{zm}}_b
\, |\widetilde{\text{ZM}}\rangle 
\ , 
\\
\Xi_3
& = 
\langle \widetilde{\text{ZM}}| \, 
\overline{\psi_a^{\text{osc}}}(t,x) \psi^{\text{zm}}_a
\overline{\psi_b^{\text{zm}}} \psi^{\text{osc}}_b(t',x')
\, |\widetilde{\text{ZM}}\rangle 
\ , 
\\
\Xi_4
& = 
\langle \widetilde{\text{ZM}}| \, 
\overline{\psi_a^{\text{zm}}} \psi^{\text{osc}}_a(t,x)
\overline{\psi_b^{\text{osc}}}(t',x') \psi^{\text{zm}}_b
\, |\widetilde{\text{ZM}}\rangle 
\ , 
\\
\Xi_5 
& = 
\langle \widetilde{\text{ZM}}| \, 
\overline{\psi_a^{\text{zm}}} \psi^{\text{zm}}_a
\overline{\psi_b^{\text{osc}}}(t',x') \psi^{\text{osc}}_b(t',x')
\, |\widetilde{\text{ZM}}\rangle 
\ , 
\\
\Xi_6
& = 
\langle \widetilde{\text{ZM}}| \, 
\overline{\psi_a^{\text{zm}}} \psi^{\text{zm}}_a
\overline{\psi_b^{\text{zm}}} \psi^{\text{zm}}_b
\, |\widetilde{\text{ZM}}\rangle 
\ , 
\end{align}
\end{subequations}
and each repeated spinor index is summed over. 

For $\Xi_1$, we may proceed as in the derivation of formula 
\eqref{AppCorrFunct-final}
in Appendix~\ref{AppenB}. Dropping the ill-defined second term in the counterpart of 
\eqref{AppCorrFunct-final}, $\Xi_1$ reduces to
$W^{(2,\overline2)}_{\text{osc}}(t,x; t', x')$ 
as evaluated in subsection~\ref{subsubsec:oscillatormodes}, 
with the result given in~\eqref{eq:W2-osc-final}. 

$\Xi_6$ reduces to $W^{(2,\overline2)}_{\text{zm}}$ 
as evaluated in subsection~\ref{subsubsec:zeromode}, 
with the result~\eqref{eq:W2-zeromode}. 

$\Xi_2$ is proportional to 
$\Tr \bigl[ S_{ab}^{\text{osc}-}(t,x; t,x) \bigr]$, 
where $S_{ab}^{\text{osc}-}$ is given by~\eqref{eq:NegWigFunct-2-osc}. 
This expression is not well defined because of the coincidence limit, 
but we interpret the expression as zero by the 
tracelessness of the gamma-matrices. 
Similarly, we interpret $\Xi_5$ as zero. 

For $\Xi_3$ and $\Xi_4$ we have
\begin{align}
\Xi_3 + \Xi_4
& = 
\langle \widetilde{\text{ZM}}| \, 
\overline{\psi_a^{\text{osc}}}(t,x) 
\psi^{\text{osc}}_b(t',x')
\psi^{\text{zm}}_a
\overline{\psi_b^{\text{zm}}} 
\, |\widetilde{\text{ZM}}\rangle 
\notag
\\
&
\hspace{3ex}
+ 
\langle \widetilde{\text{ZM}}| \, 
\psi^{\text{osc}}_a(t,x)
\overline{\psi_b^{\text{osc}}}(t',x') 
\overline{\psi_a^{\text{zm}}} 
\psi^{\text{zm}}_b
\, |\widetilde{\text{ZM}}\rangle 
\notag
\\
&= 
\langle 0^{\text{osc}}|
\overline{\psi_a^{\text{osc}}}(t,x) 
\psi^{\text{osc}}_b(t',x')
| 0^{\text{osc}} \rangle
\langle\text{ZM}|
\psi^{\text{zm}}_a
\overline{\psi_b^{\text{zm}}} 
|\text{ZM}\rangle 
\notag
\\
&
\hspace{3ex}
+ 
\langle 0^{\text{osc}}|
\psi^{\text{osc}}_a(t,x)
\overline{\psi_b^{\text{osc}}}(t',x') 
| 0^{\text{osc}} \rangle
\langle\text{ZM}|
\overline{\psi_a^{\text{zm}}} 
\psi^{\text{zm}}_b
|\text{ZM}\rangle 
\notag
\\
&= 
i \partial_\mu G^{\text{osc}+}(t,x; t',x') 
\left\{ 
\left(\gamma^\mu \right)_{ba}  
\langle\text{ZM}|
\psi^{\text{zm}}_a
\overline{\psi_b^{\text{zm}}} 
|\text{ZM}\rangle 
+ 
\left(\gamma^\mu \right)_{ab}  
\langle\text{ZM}|
\overline{\psi_a^{\text{zm}}} 
\psi^{\text{zm}}_b
|\text{ZM}\rangle 
\right\} 
\notag
\\
&= 
i \partial_\mu G^{\text{osc}+}(t,x; t',x') 
\left(\gamma^\mu \right)_{ab}  
\langle\text{ZM}|
\bigl\{\overline{\psi_a^{\text{zm}}} , 
\psi^{\text{zm}}_b \bigr\}
|\text{ZM}\rangle 
\notag
\\
&= 
i \partial_\mu G^{\text{osc}+}(t,x; t',x')  
\Tr(\gamma^0 \gamma^\mu) /L
\notag
\\
&= 
(2i/L)  \partial_t G^{\text{osc}+}(t,x; t',x') 
\notag
\\
&= L^{-2}
\sum_{n=1}^{\infty}
\left( 
e^{-2 \pi i n (\Delta t+ \Delta x -i\epsilon)/L} 
+ e^{-2 \pi i n (\Delta t- \Delta x -i\epsilon)/L} 
\right) 
\notag
\\
&= L^{-2}
\left( 
\frac{1}{e^{2 \pi i(\Delta t+ \Delta x -i\epsilon)/L} -1}
+ \frac{1}{e^{2 \pi i(\Delta t- \Delta x -i\epsilon)/L} -1}
\right)
\ , 
\end{align}
using \eqref{eq:bothWigFunct-2-osc}, \eqref{eq:Gplus-cylinder-osc-sum} 
and~\eqref{eq:cylinder-et-zm-anticomm}. 

Collecting these results yields~\eqref{eq:untwisted-totalW2}.

\section{Stationary detector in $(1+1)$ Minkowski vacuum 
with Gaussian switching\label{app:stationary1+1}}

In this appendix we evaluate the response of an inertial 
detector and a uniformly accelerated detector 
in $(1+1)$ Minkowski vacuum with a Gaussian switching. 

\subsection{Inertial detector}

Inserting the inertial worldline \eqref{eq:inertial-worldline} 
and the Gaussian switching \eqref{eq:gaussian-switching} in 
\eqref{Mink-RespFunct-massless-final} with $d=2$, we may change variables by 
$\tau-\tau' = \sigma z$ and $\tau+\tau' = \sigma w$ 
and perform the Gaussian integral over~$w$, with the result 
\begin{align}
\mathcal{F}(\Omega) = 
- \frac{1}{2\pi^2 \sigma} H(\sigma\Omega)
\ , 
\end{align}
where the function $H$ of a real variable is defined by 
\begin{align}
H(\alpha) := 
\int_C dz \, \frac{\exp(-i\alpha z - z^2/4)}{z^2} 
\ , 
\label{eq:H-def}
\end{align}
where the contour $C$ follows the real axis from $-\infty$ to $\infty$ 
except for dipping into the lower half-plane near $z=0$. 
Differentiating \eqref{eq:H-def} twice and evaluating the Gaussian integral gives 
$H''(\alpha) = - 2 \pi^{1/2} e^{-\alpha^2}$, and integrating this twice gives 
\begin{align}
H(\alpha) = \pi \left( \alpha \erfc\alpha - \frac{e^{-\alpha^2}}{\pi^{1/2}} \right) 
+ A\alpha + B
\ , 
\label{eq:H-erfc-w-AB}
\end{align}
where $\erfc$ is the error complement function \cite{dlmf} 
and $A$ and $B$ are constants. 

To determine $A$ and $B$, we deform the contour $C$ in \eqref{eq:H-def} to 
$z = u - i$ with $u\in\BbbR$, which gives the estimate 
\begin{align}
|H(\alpha)| \le  
e^{- \alpha + 1/4}
\int_{-\infty}^{\infty} du  \, \frac{\exp(- u^2/4)}{u^2+1} 
\ , 
\label{eq:H-alt}
\end{align}
which shows that $H(\alpha) \to 0$ as $\alpha\to\infty$. 
The falloff of $\erfc$ at large positive argument 
then shows that $A=B=0$ in~\eqref{eq:H-erfc-w-AB}. 

Collecting, 
\begin{align}
\mathcal{F}(\Omega) = 
\frac{1}{2\pi \sigma} 
\left( 
\frac{e^{-\sigma^2\Omega^2}}{\pi^{1/2}}
- 
\sigma\Omega \erfc(\sigma\Omega) 
\right) 
\ . 
\end{align}

\subsection{Uniformly accelerated detector}

Substituting the uniformly accelerated trajectory \eqref{eq:rindler-trajectory} 
and the Gaussian switching \eqref{eq:chi-gauss-taunought} in  
\eqref{Mink-RespFunct-massless-final} with $d=2$, we change variables by $a(\tau-\tau') = 2z$ 
and $a(\tau+\tau'-2\tau_0) = 2w$ and perform the Gaussian integral over~$w$, with the result 
\begin{align}
\mathcal{F}(\Omega) 
&= 
- \frac{a}{4\pi^2}
\int_{-\infty}^\infty
\frac{dz}{\sinh^2(z-i\epsilon)}
\, 
\exp \! \left( -\frac{z^2}{\sigma^2a^2}
-\frac{2i\Omega z}{a} \right)
\notag
\\
&= 
\frac{a e^{-\pi\Omega/a}}{4\pi^2}
\int_{-\infty}^\infty
\frac{dz}{\cosh^{2\!} r}
\, 
\exp \! \left( -\frac{{(r - i \pi/2)}^2}{\sigma^2a^2}
-\frac{2i\Omega r}{a} \right)
\ , 
% \label{eq:erl11}
\end{align}
where in the last equality we have deformed the contour to $z = r - i\pi/2$ with $r\in\BbbR$.

\end{document}